\documentclass[twocolumn,aps,prd,amsmath,amssymb,preprintnumbers,%
longbibliography]{revtex4-1}

\makeatletter							
\@ifclasswith{revtex4-1}{draft}{		
	\usepackage[notref,notcite]{showkeys}
	
}{}
\makeatother

\usepackage{amsmath}
\usepackage{amssymb}
\usepackage{bbm}
\usepackage{epsfig}
\usepackage{graphics}
\usepackage{graphicx}
\usepackage{titlesec}
\usepackage{mathtools}
\usepackage{environ}
\usepackage{version}
\usepackage[colorlinks=true]{hyperref}
\hypersetup{urlcolor=black,linkcolor=black,citecolor=black}
\usepackage{natbib}
\usepackage[toc,page]{appendix}

\usepackage{enumerate}

\usepackage{dsfont}

\usepackage{caption}
\usepackage{subcaption}
\usepackage{tabularx}
\newcolumntype{C}[1]{>{\centering\arraybackslash}p{#1}}

\makeatletter
\renewcommand*\env@matrix[1][c]{\hskip -\arraycolsep
 \let\@ifnextchar\new@ifnextchar
 \array{*\c@MaxMatrixCols #1}}
\makeatother

\newcommand{\be}{\begin{equation}}
\newcommand{\ee}{\end{equation}}
\newcommand{\bel}[1]{\be\label{#1}}

\newcommand{\ba}{\begin{align}}
\newcommand{\ea}{\end{align}}
\newcommand{\nn}{\nonumber}

\newcommand{\gl}{\big(}
\newcommand{\gr}{\big)}

\titleformat{\subsection}[block]{\normalfont\bfseries}{\thesubsection.}{1ex}{}
\titlespacing{\subsection}{0pt}{10pt}{1pt}[0pt]
\titleformat*{\section}{\large\bfseries}
\renewcommand{\thesubsection}{\arabic{subsection}}

\pdfoutput=1 

\usepackage{todonotes}
\usepackage{braket}

\newcommand{\tr}{\mathrm{tr}}

\newcommand{\psibar}{\overline{\psi}}

\newcommand{\cD}{\mathcal{D}}
\newcommand{\cR}{\mathcal{R}}

\newcommand{\cA}{\mathcal{A}}

\newcommand{\vptil}{\tilde{\vp}}

\newcommand{\ztil}{\tilde{z}}

\newcommand{\Shat}{\widehat{S}}
\newcommand{\vp}{{\varphi}}

\newcommand{\eps}{\varepsilon}

\newcommand{\Sbar}{\bar{S}}

\newcommand{\chat}{\hat{\chi}}

\newcommand{\qq}[1]{``#1''}

\definecolor{refkey}{rgb}{0,0,1}
\definecolor{labelkey}{rgb}{0,1,0}

\begin{document}

\title{\LARGE Field transformations in functional integral, effective action and
functional flow equations}

\author{C. Wetterich}

\affiliation{Institut f\"ur Theoretische Physik\\
Universit\"at Heidelberg\\
Philosophenweg 16, D-69120 Heidelberg}

\begin{abstract}

Field transformations for the quantum effective action lead to different
pictures of a given physical situation, as describing a given evolution of the
universe by different geometries. Field transformations for functional flow
equations can reveal important physical features, as the appearance of bound
states. They also allow for technical simplifications. We make a critical
assessment of the virtues and shortcomings of different versions of field
transformations. Key issues are demonstrated by the quantum field theory for
scalars with a field-dependent coefficient of the kinetic term. Our findings
confirm the principle of field relativity for gravity and cosmology based on the
quantum effective action.

\end{abstract}

\maketitle

\section{Introduction}\label{sec:I}

Field transformations are widely used in many areas of physics. In the presence
of fluctuations -- quantum or thermal -- it matters for which object the
transformations are performed. For example, in quantum gravity a field
transformation of the classical action yields different results as the
transformation of the quantum effective action. This difference results in the
role of fluctuations. A proper field transformation of the functional integral
has to include the transformation of the functional measure. The corresponding
Jacobian is often not taken into account, leading to statements that theories
related by field transformations are not equivalent on the quantum level. In
cosmology and quantum gravity the omission of the Jacobian has led to a long
debate on the equivalence of models related by field transformation, see the
review~\cite{FAGU}. In condensed matter physics this issue concerns the
difference of making field transformations in the Hamiltonian or the free
energy. They are only equivalent if the transformation of the measure factor is
taken into account.

In the presence of fluctuations the relevant field equations are the ones
obtained by variation of the quantum effective action or free energy with
respect to the fields. They are exact identities for a given form of the
effective action and contain the relevant physical information, in contrast to
the field equations derived by variation of the classical action. Field
transformations of the effective action, that we call here \qq{macroscopic field
transformations}, lead to identical field equations, simply related by a
variable transformations in differential equations. The quantum effective action
or free energy contains much more information than only the field equations. The
second functional derivative is related to the connected correlation function or
inverse propagator, and interaction vertices follow from higher functional
derivatives. After a field transformation the higher functional derivatives of
the effective action may no longer be directly related to microscopic
correlation functions. They still play an important role for macroscopic
observations which are often not sensitive to the microscopic details or
microscopic correlation functions. A well known example is the universality of
critical phenomena where vastly different microscopic theories can lead to the
same free energy or effective action. The second functional derivative of the
effective action contains the information about the correlation length
independently of the microscopic model. We name the functional derivatives of
the effective action \qq{macroscopic correlation functions} and discuss how
their relation to microscopic correlation functions is affected by field
transformations.

This paper aims for an overview of the impact of field transformations on
different levels. In the view of recent discussions we put particular emphasis
on the technical simplifications that can be realized by field transformations
in functional flow equations.

Field transformations can help to focus on important physical features and to
discard redundant information. For critical phenomena in statistical physics
their use has a long tradition~\cite{WEG}. Various versions of field
transformations have been formulated for modern functional flow
equations~\cite{CWFT, GW1, GW2, PAW, FW}. They are used for making important
\qq{composite or collective degrees of freedom} of a model directly accessible
or for technical simplifications. Motivated by the elimination of redundant
couplings for the counting of relevant parameters at a given fixed point of
renormalization group equations~\cite{WEG}, the proposal of the \qq{essential
renormalization group}~\cite{BZF, BAFA} has argued that couplings which can be
removed by field transformations are not observable and need not to be
considered for the search of possible scaling solutions. We question this
statement. The essential renormalization group entails important technical
simplifications for approximate solutions of functional flow
equations~\cite{BKN1, KPRS, LLPS, IPAW2, BPER, KPS, BKS, SAU, IPA, KO, BKN2,
BFK}. A critical discussion of the merits and shortcomings of this approach
seems timely.

Removing couplings by field transformations can be of great help but also
entails substantial dangers. This may be demonstrated by two cases. The
introduction of renormalized fields removes a field-independent and
momentum-independent wave function renormalization $Z$. This quantity is not
observable for purely macroscopic observations. The wave function
renormalization only plays a role if macroscopic observables are related to
microphysical degrees of freedom. For example, it is crucial for
understanding~\cite{CWAPT, GRCW, GEW} the compatibility of the
Kosterlitz-Thouless phase transition~\cite{KT} with the Mermin-Wagner
theorem~\cite{MW}. If $Z$ depends on some renormalization scale $k$, its
logarithmic $k$-derivative, $\eta=-k\partial_k\ln Z$, defines an anomalous
dimension. This appears in renormalization group or flow equations once $Z$ is
removed by a field transformation. The same anomalous dimension often governs
the details of the momentum- or field-dependence of the propagator, as encoded
by the replacement of $Z$ by a momentum- and field-dependent factor $Z(q,\chi)$
multiplying the kinetic term.

If the dependence of $Z(q,\chi)$ on the momenta $q$ and fields $\chi$ is
structureless and the anomalous dimension is small, it seems a promising idea to
absorb $Z(q,\chi)$ by a field transformation. This simplification of the
propagator facilitates many computations. In particular, for a field-independent
$Z(q)$ a linear field transformation moves the information contained in $Z(q)$
to \qq{dressed vertices} by multiplying the vertices with appropriate powers of
$Z^{-1/2}(q)$. The dressed vertices are the same for all original choices of
fields which are related by a linear momentum-dependent field transformation. In
this sense the dressed vertices contain the \qq{physical} information.

Our second example points to the opposite direction. The inverse propagator of
electrons in a solid is of the form $P_F(q)=\eps(q)-\mu+i(2n+1)\pi T$, where
$\mu$ is the chemical potential, $T$ the temperature and the integer $n$
determines the Matsubara frequency. The Fermi surface is determined by
$\eps(q_F)=\mu$. It is located at non-zero Fermi momenta $q_F$ even for the
simplest case $\eps(q)=q^2/(2m)$. In solids $\eps(q)$ can have a rather complex
form which fixes the detailed location of the Fermi surface and many other
directly observable features. For $T\neq0$ a regular linear field transformation
transforms $P_F(q)\to\tilde P_F(q)=f(q)P_F(q)$. We can bring $\tilde P_F(q)$ to
an arbitrarily chosen form without zeros. After the field transformation $\tilde
P_F(q)$ does no longer contain the information about the Fermi surface if we
take some standard form as $\tilde P_F=q^2/(2m)+i(2n+1)\pi T$ for a gas of free
fermions or bosons at zero density. The corresponding complex function $f(q)$
has removed the chemical potential and all structures in the dispersion
relation $\eps(q)$.

The multiplication of the fermion fields by $f^{-1/2}(q)$ is linear, but highly
non-local. Powers of $f^{1/2}(q)$ multiply the vertices once expressed in terms
of the rescaled fields. This includes the source term $\int_qj(q)\psi(q) =
\int_q\tilde j(q)\tilde \psi(q)$, $\tilde j(q) = f^{1/2}(q)j(q)$. If the source
term is local in the original fermion field $\psi(q)$, it will be non-local for
the transformed field $\tilde\psi(q)$. The source term is not directly visible
in the effective action. Deriving field equations for non-zero sources like
the non-local transformation transformation of the source becomes important.

The details of the dispersion relation involves couplings that parameterize the
electron propagator in a solid. For the two-dimensional Hubbard model with
next-neighbor and next-to-nearest neighbor hopping parameters $t$ and $t'$ one
has
\bel{AX0}
\eps(q)=-2t(\cos q_x+\cos q_y)-4t'\cos q_x\cos q_y\ .
\ee
One of the couplings that is removed by the above field transformation is the
ratio $t'/t$. The conclusion that couplings used to parameterize the
momentum-dependence of the fermion propagator are unobservable is not justified
even for vanishing sources. The physical information contained in these
couplings is moved by the field transformation to the vertices. Only certain
products of vertices with powers of $P_F^{-1/2}(q)$ are invariant under the
momentum-dependent field transformation. If one would use a truncation which
neglects the detailed momentum-dependence of vertices, key physical information
would even be lost by a field transformation which brings the electron
propagator to some standard form.

The danger of discarding important information, which is already visible for
linear field transformations, gets greatly enhanced for non-linear field
transformations. The key issue is the control to which sector the information
about observables is moved by a field transformation. Choosing optimal fields
for the solution of a given problem is mainly a question of simplicity. Since
truncations of the effective action are usually needed, it is crucial to encode
the most relevant physical information in the part that is not discarded by the
truncation. If important physical features are visible in the propagator, it
seems to be a good idea to keep as much information as possible in the field-
and momentum-dependent propagator, extended to low order vertices with perhaps
less complete resolution.

Different versions of field transformations for functional flow equations have
been proposed in the past. These functional flow equations describe the
dependence of the effective average action $\Gamma_k$ on a renormalization scale
$k$ which typically corresponds to an infrared cutoff. Only fluctuations with
momenta greater than $k$ are then included for $\Gamma_k$. For $k\to0$ the
effective average action becomes the usual effective action with all quantum or
thermal fluctuations included. The present paper tries to compare the advantages
and disadvantages of main representatives of field transformations in functional
flow equations. Particular emphasis is devoted to the issue of control of the
field transformation. We also address the question to which extent the effective
actions obtained by different versions of field transformations are equivalent.
Furthermore, we assess the relation between macroscopic correlation functions
obtained from functional derivatives of the effective action and microscopic
correlation functions as defined by expectation values of powers of microscopic
fields.

The possibility of field transformations has also important conceptual
implications. This is most striking for gravity. A non-linear transformation of
the metric field changes the geometry. For example, an expanding universe can be
transformed into a shrinking universe~\cite{CWUE, BST, CWHBSF, CWET, MLI, HMSS,
DS1, IJS, CP, DS2, KPV, GHR, AB, MSH, LQ, CKK, KZ, RUW, MUS, LL}. Field
relativity~\cite{CWUE} states that macroscopic field transformations of the
quantum effective action do not change the observables. Cosmologies with
different geometries related by such transformations are strictly equivalent.
Of course, this necessitates the simultaneous transformation of the energy
momentum tensor as a source term~\cite{CWVNC}. The present paper confirms this
point of view. Field relativity also establishes the equivalence of many models
of modified gravity with coupled quintessence~\cite{CWMG}.

As a simple example illustrating various aspects of field transformations we
focus in this paper on a quantum field theory for a single scalar field, with a
classical action involving only a term with two derivatives, multiplied by a
field-dependent wave function renormalization or \qq{kinetial}. On the one hand
a non-linear field transformation can transform away the field-dependence of the
kinetial such that the transformed action looks like the one for a free massless
scalar field. On the other hand the field-dependence of the kinetial leads to
terms in the action involving more than two powers of the field, suggesting
interactions. We will relate these various aspects and find that for generic
kinetials this model indeed describes an interacting scalar field.

While a general field-dependent kinetial describes a theory with interactions,
we find strong indications for an ultraviolet fixed point for which the
effective average action is given for $\chi\to0$ by
\bel{eq:AA}
\Gamma_k[\chi]=\int_x\left(\frac{\kappa_0}{2\chi^2}\partial^\mu\chi\partial_\mu
\chi+\text{const.}\right)\ .
\ee
This describes a free massless composite field $\vp\sim\ln\chi$. This
ultraviolet fixed point differs from the (trivial) infrared fixed point for
which the kinetial is field-independent. The ultraviolet fixed point is
characterized by the quantum scale symmetry of a constant multiplicative
rescaling $\chi\to\beta\chi$ with fixed metric. In contrast, the infrared fixed
point exhibits a different form of quantum scale symmetry for which the metric
(or coordinates) are rescaled together with $\chi$. The effective action for the
infrared fixed point is invariant under constant shifts $\chi\to\chi+\delta$.
Possible crossover trajectories between the two fixed points are characterized
by the presence of interactions. If the ultraviolet fixed point~\eqref{eq:AA} is
confirmed this would define a possible ultraviolet completion of an interacting
scalar field theory in four dimensions, avoiding the \qq{triviality problem}.

In sect.~\ref{sec:FTFI} we begin with field-transformations for the integration
variables of the functional integral defining the quantum field theory. The
Jacobian of this transformation modifies the action beyond a simple replacement
of the field variables. We discuss how field transformations affect the relation
between macroscopic and microscopic correlation functions. On the level of the
effective action we demonstrate how non-linear field transformations change the
saddle point approximation and perturbation theory. Sect.~\ref{sec:FTEA} turns
to field transformations for the quantum effective action for which all effects
of quantum fluctuations are included. We distinguish between macroscopic field
transformations which only modify the expression of a given effective action in
terms of field-variables, and a type of microscopic field transformations which
modify in addition the sources used for the definition of the effective action.

In sect.~\ref{sec:FTFE} we proceed to the field transformations in functional
flow equations, focusing first on the macroscopic field
transformations~\cite{CWFT, GW1}. This is contrasted in sect.~\ref{sec:FECS}
with the microscopic field transformations~\cite{PAW, BZF, BAFA, IPA} which are
the basis of the proposed \qq{essential renormalization group}. In
sect.~\ref{sec:TFF} we employ a formulation with both fundamental and composite
fields in order to clarify further aspects of the microscopic field
transformations, in particular the relation between the classical action and the
initial values of the flow equations. Sect.~\ref{sec:SS} concentrates on scaling
solutions of the flow equations whose existence can make a model ultraviolet
complete. A conclusding discussion is presented in sect.~\ref{sec:C}. Two
appendices provide a discussion of the model with a field-dependent kinetial
without the use of field transformations. This can be used for comparison with
treatments involving field transformations. These computations yield indications
for the exitence of the ultraviolet fixed point~\eqref{eq:AA}.

\section{Field transformations for functional\\integrals}\label{sec:FTFI}

Field transformations in a functional integral are conceptually very simple.
They are the analogue of variable transformations in ordinary integrals. The
Jacobian of the transformation plays an important role and is often difficult to
handle in practice. We demonstrate this for a simple model of a scalar field
with a field-dependent kinetic term. The classical action involves no potential
or mass term, while the coefficient of the term with two derivatives - the
kinetial - is allowed to depend on the value of the scalar field. A field
transformation can transform this classical action to the one for a free
massless scalar field. The Jacobian induces, however, additional terms in the
relevant microscopic action. By virtue of the Jacobian the interactions
contained in the kinetial are shifted to a potential.

\subsection*{Functional integral}

Let us define our model by the \qq{classical} or \qq{microscopic} action
\bel{M1}
S[\chi']=\frac12\int_xK(\chi')\partial^\mu\chi'\partial_\mu\chi'\ ,
\ee
with $\partial^\mu\chi'=\eta^{\mu\nu}\partial_\nu\chi'$, $\eta^{\mu\nu}$ the
metric of flat Euclidean or Minkowski space, $\int_x$ the integral over
$d$-dimensional space $\mathbb{R}^d$ with Cartesian coordinates $x^\mu$, and
$\partial_\mu\chi'=\partial\chi'/\partial x^\mu$. For Euclidean signature the
action defines a probability density $p[\chi']$ for field configurations
$\chi'(x)$,
\bel{M2}
p[\chi']=Z^{-1}\exp\gl-S[\chi']\gr\ ,
\ee
with partition function given by the functional integral
\bel{M3}
Z=\int\cD\chi'\,\exp\gl-S[\chi']\gr\ .
\ee
The functional measure is formally defined by
\bel{M4}
\int\cD\chi'=\prod_x\int_{-\infty}^\infty\text{d}\chi'(x)\ .
\ee
This formal definition may be considered as the continuum limit of some lattice
regularization. A field $\chi'$ with linear measure~\eqref{M4} and action
expressed as functional of $\chi'$ will later be called a \qq{canonical field}.

One is typically interested in correlation functions as
\bel{M5}
G(x,y)=\langle\chi'(x)\chi'(y)\rangle-\langle\chi'(x)\rangle
\langle\chi'(y)\rangle\ ,
\ee
or similar higher order correlation functions. As usual, the expectation value
for an observable is given by
\bel{A5}
\langle A\rangle=Z^{-1}\int\cD\chi'A[\chi']\exp\gl-S[\chi']\gr\ .
\ee
(We will later generalize this expression to the presence of sources.)
If the \qq{kinetial} $K(\chi')$ is not constant, the model involves
interactions, as manifest in the classical field equation obtained by variation
of the action
\bel{M6}
K\partial^\mu\partial_\mu\chi'=-\frac12\frac{\partial
K}{\partial\chi'}\partial^\mu\chi'\partial_\mu\chi'\ .
\ee
We will see that these interactions can lead to complex phenomena as spontaneous
symmetry breaking.

\subsection*{Field transformations}

By a non-linear field transformation
\bel{M7}
\vp'(x)=\vp'\gl\chi'(x)\gr\ ,
\ee
defined by
\bel{M8}
\frac{\partial\vp'}{\partial\chi'}=K^{1/2}(\chi')\ ,
\ee
one can bring the classical action~\eqref{M1} to the form of a free scalar field
theory
\bel{M9}
S=\frac12\int_x\partial^\mu\vp'\partial_\mu\vp'\ .
\ee
This is, however, not the relevant action for the scalar field $\vp'$. The
non-linear field transformation involves a Jacobian in the functional integral
\begin{align}
\label{M10}
Z=&\,\int\cD\chi'\,e^{-S[\chi']}=\int\cD\vp'\,J[\vp']
e^{-S\big[\chi'(\vp')\big]}\nn\\ =&\,\int\cD\vp'\,e^{-\gl S[\vp']-\ln
J[\vp']\gr}=\int\cD\vp'\,e^{-\Sbar[\vp']}\ ,
\end{align}
with
\bel{M11}
J[\vp']=\prod_x\left|\frac{\partial\chi'(x)}{\partial\vp'(x)}\right|=
\exp\left(-\eps^{-d}\int_x\ln\left|\frac{\partial\vp'}{\partial\chi'}
\right|\right)\ .
\ee
Here $\eps$ is a typical length scale (lattice distance) related to the
regularization, defined by $\sum_x=\eps^{-d}\int_x$.

As a consequence, the action for $\vp'$ involves a potential term reflecting the
interactions
\bel{M12}
\Sbar[\vp']=\int_x\left\{\frac12\partial^\mu\vp'
\partial_\mu\vp'+V(\vp')\right\}\ ,
\ee
where
\bel{M13}
V(\vp')=\frac{\eps^{-d}}{2}\ln K[\vp']\ .
\ee
We suppose that $K[\chi']$ is positive for all $\chi'$, such that taking the
positive root in eq.~\eqref{M8} results in $\vp'$ monotonically increasing with
$\chi'$.

Inversely, many scalar theories with a potential and action~\eqref{M12} can be
cast by a non-linear field transformation into the form~\eqref{M1}, with
kinetial
\bel{M14}
K=\exp\gl2\eps^{d}V\gr\ .
\ee
After the field transformation $\vp'\to\chi'(\vp')$ the potential is no longer
present, $V(\chi')=0$. It has been absorbed by the Jacobian. As an example we
may take
\bel{M15}
V=\frac{\eps^{-d}}{2\lambda}\mathrm{tanh}^2\left\{
\frac{\lambda\eps^{d/2}}{2}\gl\vp^{\prime2}-\vp_0^2\gr\right\}\ ,
\ee
where $\vp_0^2$ is a positive or negative constant. For small
$|\vp^{\prime2}-\vp_0^2|\ll\eps^{-d/2}/\lambda$ we can expand
\bel{M16}
V=\frac\lambda8\gl\vp^{\prime2}-\vp_0^2\gr^2+\hdots
\ee
For $\vp_0^2>0$ and small $\lambda>0$ this is a typical model with spontaneous
symmetry breaking in the perturbative range. The kinetial~\eqref{M14} reads
\bel{M17}
K=\exp\left\{\frac1\lambda\mathrm{tanh}^2\left[\frac{\lambda\eps^{d/2}}{2}\gl\vp^{\prime2}-\vp_0^2\gr\right]\right\}\
.
\ee
If the potential $V(\vp')$ in a model with canonical kinetic term leads to
spontaneous symmetry breaking, the same holds after the field transformation in
the model without potential, but with a non-trivial kinetial. This is a simple
demonstration that a non-trivial kinetial can lead to complex phenomena and
should not be \qq{discussed away}.

For $\vp^{\prime2}$ near $\vp_0^2$ one has $K\approx1$ such that $\chi'$ and
$\vp'$ almost coincide. On the other hand, for $\vp^{\prime2}\to\infty$ one
infers $K\approx\exp(1/\lambda)$ and therefore
\bel{M18}
\chi'\approx e^{-\frac1{2\lambda}}\vp'+\bar\chi\ ,
\ee
where the constant $\bar\chi$ can be found by numerical integration of
eq.~\eqref{M8}. We choose $\chi'(\vp'=0)=0$ such that
$\chi'(-\vp')=-\chi'(\vp')$ and $K_0=K(\chi'=0)\approx1+\lambda\eps^d\vp_0^4/4$.
For $\vp_0^2>0$ and $\chi'_0=\chi'(\vp_0)$ one has $K(\chi'_0)=1$. In this case
$K(\chi')$ interpolates smoothly between $K(\chi'=0)=K_0$, $K(\chi'_0)=1$ and
$K\gl|\chi'|\to\infty\gr\approx\exp(1/\lambda)$. A minimum of $K(\chi')$ at
$\chi'_0\neq0$ results in a minimum of $V(\vp')$ at $\vp_0=\vp'(\chi'_0)$.
Nothing prevents us to make this choice of $K(\chi')$. Our simple example
demonstrates that large classes of scalar theories, including ones with
non-trivial features as spontaneous symmetry breaking, can be described by the
classical action~\eqref{M1} for a suitable kinetial $K(\chi')$.

The partition function does not depend on the choice of the integration variable
$\chi'$ or $\vp'$, provided one accounts properly for the Jacobian,
\bel{M19}
Z=\int\cD\chi'\,\exp\gl-S[\chi']\gr=\int\cD\vp'\,\exp\gl-\Sbar[\vp']\gr\ .
\ee
Also the probability distribution (for Euclidean signature) is the same
\bel{M20}
Z^{-1}\cD\chi'\,\exp\gl-S[\chi']\gr=Z^{-1}\cD\vp'\,\exp\gl-\Sbar[\vp']\gr\ .
\ee
One concludes that all correlation functions are identical for both
formulations, provided they are expressed properly in terms of $\chi'$ or
$\vp'$. For example, the connected two-point function~\eqref{M5} involves
non-linear expressions in $\vp'$ once $\chi'(\vp')$ is inserted. Simple
correlation functions for one choice of fields may look rather complicated for
another choice of fields. Nevertheless, a possible expectation value
$\langle\chi'(x)\rangle=\chi'_0$ translates directly to an expectation value
$\langle\vp'(x)\rangle=\vp'(\chi'_0)$.

\subsection*{Saddle point approximation}

In principle, it does not matter which fields are chosen for the functional
integral. This does no longer hold for given approximations. We demonstrate this
for the saddle point approximation for our model. In the tree approximation one
evaluates $\langle\vp'\rangle$ by a solution of the field equation derived from
the classical action $\Sbar[\vp']$. For our example with $\vp_0^2>0$ this
results in $\langle\vp'(x)\rangle=\pm\vp_0$, as given by the minimum of the
potential $V(\vp')$. In contrast, the tree approximation for
$\langle\chi'\rangle$, defined by a solution of the field equation~\eqref{M6}
derived by variation of $S[\chi']$, admits arbitrary constant
$\langle\chi'(x)\rangle=\text{const.}$ The tree approximation in the two
pictures looks rather different. This is not surprising, since the tree
approximation does not account for the role of the Jacobian appearing in the
treatment of fluctuations. Performing a loop expansion in both schemes the
difference in the loop contributions has to match the difference in the tree
approximations such that the result of $\langle\vp'(x)\rangle$ is identical in
both schemes.

Let us first consider a saddle point approximation in the $\chi$-scheme with
action~\eqref{M1}. In one-loop order the effective action reads (see below)
\bel{M21}
\Gamma[\chi]=S[\chi]+\frac12\tr\big\{\ln S^{(2)}[\chi]\big\}\ ,
\ee
with $S^{(2)}$ the second functional derivative of $S$,
\begin{align}
\label{M22}
S^{(2)}(x,y)=-\Big[&\,K\partial^\mu\partial_\mu+\partial_\chi
K\partial^\mu\chi\partial_\mu+\frac12\partial^2_\chi
K\partial^\mu\chi\partial_\mu\chi\nn\\
&\,+\partial_\chi K\partial^\mu\partial_\mu\chi\Big](x)\delta(x-y)\ .
\end{align}
Here all derivatives are taken with respect to $x$ and $\partial_\chi K=\partial
K/\partial\chi$. The argument $\chi(x)$ of the effective action is the
\qq{macroscopic} or \qq{background} field which can be identified with the
expectation value $\langle\chi'(x)\rangle$ in the presence of suitable sources.

The potential term $U(\chi)$ in the effective action obtains by evaluating
eq.~\eqref{M21} for a constant background field,
\bel{M23}
\Gamma[\chi]=\int_xU(\chi)=\frac12\tr\big\{\ln
K(\chi)+\ln\gl-\partial^\mu\partial_\mu\gr\big\}\ .
\ee
For the first term we evaluate the trace in position space,
\bel{M24}
\tr\ln K=\sum_x\ln K=\eps^{-d}\int_x\ln K\ ,
\ee
and we extract
\bel{M25}
U(\chi)=\frac{\eps^{-d}}{2}\ln K(\chi)\ .
\ee
This is precisely the potential~\eqref{M13} which is present in the tree
approximation in the $\vp$-scheme. The trace for the second term may be
evaluated in momentum space where $-\partial^\mu\partial_\mu$ translates to the
squared momentum $q^2=q^\mu q_\mu$. The second term yields a field-independent
constant which plays no role.

For a general $x$-dependent field $\chi(x)$ we write
\bel{M26}
\tr\big\{\ln S^{(2)}\big\}=\tr\ln K+\tr\ln\Delta\ ,
\ee
with
\bel{M27}
\Delta=K^{-1}S^{(2)}=-\gl\partial^\mu\partial_\mu+B^\mu\partial_\mu+C\gr\ ,
\ee
where
\bel{M28}
B^\mu=\partial_\chi\ln K\partial^\mu\chi\ ,
\ee
and
\bel{M29}
C=\frac1{2K}\partial_\chi^2K\partial^\mu\chi\partial_\mu\chi+\partial_\chi\ln
K\partial^\mu\partial_\mu\chi\ .
\ee
The first term in eq.~\eqref{M26} yields the potential~\eqref{M25}, while the
second term contributes derivative terms to the effective action.

This correction to the kinetic terms needs a regularization, as can be seen by
evaluating traces of functions of $\Delta$ in the heat kernel expansion.
Expanding in the number of derivatives one finds
\bel{M30}
\tr
f(\Delta)=\frac1{16\pi^2}\int_x\int_0^\infty\text{d}zf(z)\gl(z+H(\chi)\partial^\mu\chi\partial_\mu\chi+\hdots\gr\
.
\ee
Here $H(\chi)$ is defined by
\bel{M31b}
\int_xH(\chi)\partial^\mu\chi\partial_\mu\chi=\int_x\gl C-\frac14B^\mu B_\mu\gr\
,
\ee
which yields
\bel{M31}
H(\chi)=\frac34\gl\partial_\chi\ln K\gr^2-\frac1{2K}\partial_\chi^2K\ .
\ee
The dots in eq.~\eqref{M30} stand for terms with more than two derivatives. For
$f(z)=\ln(z)$ the $z$-integral does not converge. Even without specifying a
regularization it is clear, however, that the term $\sim H(\chi)$ results in a
renormalization of the kinetial, $K\to K+cH$. This does not affect the minimum
of the potential which determines the expectation value
$\langle\chi'(x)\rangle=\chi_0$. We conclude that the expectation value of the
scalar field is the same for the tree approximation in the $\vp$-scheme and the
one-loop approximation in the $\chi$-scheme.

Going further by including the one-loop correction in the $\vp$-scheme one finds
a renormalization of the effective scalar potential. (The one-loop correction
needs again a regularization.) This modifies the expectation value
$\langle\vp'(x)\rangle$. In turn, this change has to be matched by higher loop
terms in the $\chi$-scheme. We conclude that a change of variables in the
functional integral is allowed, in principle. In practice, it may be rather
cumbersome. This holds even if the Jacobian is computable as for our case.

We may consider the particular case
\bel{33A}
K=\frac{\kappa_0}{\chi'^2}\ .
\ee
In this case $H(\chi)$ vanishes. On the other hand, one obtains in leading order
of the saddle point expansion a potential which diverges for $\chi\to0$
\bel{33B}
U=-\eps_d\gl\ln(\chi)-\frac12\ln\kappa_0\gr\ .
\ee
We conclude that the possible ultraviolet fixed point~\eqref{eq:AA} is
unlikely to arise from a microscopic action~\eqref{M1} with linear
measure~\eqref{M4}. It could rather be connected to a measure
$\sim\text{d}\chi'/\chi'$ which respects the multiplicative scale symmetry and
is linear in $\vp'$. In this case one deals with a free massless scalar field
$\vp=\sqrt{\kappa_0}\ln(\chi/\sqrt{\kappa_0})$. For the possible ultraviolet
fixed point this should describe the behavior for $\chi\to0$, $\vp\to-\infty$. A
possible crossover to the trivial fixed point for $\chi$ needs an instability
towards $\chi\to\infty$, $\vp\to\infty$. A possibility is a measure which is
linear in $\chi'$ for large $|\chi'|$, and $\sim\text{d}\chi'/\chi'$ for small
$|\chi'|$.

\section{Field transformations for the effective action}\label{sec:FTEA}

The quantum effective action $\Gamma$ is a very strong concept in quantum field
theory. Since it incorporates all effects of quantum fluctuations. The analogue
in equilibrium statistical physics is a field-dependent Gibbs free energy which
includes all effects of thermal fluctuations. We will use the common wording
\qq{effective action} for both, as well as for similar objects for which
fluctuation effects of a functional integral or probability distribution are
fully included. The macroscopic field equations obtain by setting the first
functional derivative of the effective action equal to a source. They are exact
identities without any further corrections. For macroscopic observations these
are the only relevant \qq{classical field equations} in the context of
fluctuating microphysics. For gravity the source is the energy momentum tensor,
for electrodynamics in medium an external charge distribution, or for
superconductivity an external magnetic field.

For a given solution of the field equation one may investigate a small
perturbation of the source. The response of the solution to this perturbation in
found by linearizing the field equation around the given solution for the
unperturbed source. This involves the second functional derivative
$\Gamma^{(2)}$ of the effective action. The inverse of $\Gamma^{(2)}$ defines
the propagator as well known for the general setting of Green's functions. In
our context we call this propagator a \qq{macroscopic correlation function}.
Higher functional derivatives of $\Gamma$ define vertices and the associated
coupling constants for effective interactions.

Macroscopic observables are expressed in terms of functional derivatives of the
effective action. This relation is typically formulated for a given choice of
fields. It may be rather complex, as the extraction of the scattering matrix
from the effective action. For our purposes it is sufficient that all relevant
macroscopic observables can be computed in principle as functions of the
functional derivatives of the effective action. The effective action contains
then the full information about all relevant physical observables.

The effective action will not be known exactly except for special cases as free
field theories. Nevertheless, powerful approximations (\qq{truncations}) exist
based on symmetries and expansions in the number of derivatives, small couplings
etc. Quite often one simply assumes a valid approximation for the effective
action. This is the case for Landau type theories for superconductors or
superfluids. From this ansatz for the effective action many physical observables
are computed. For example, the macroscopic two-point function or propagator is
extracted from the second functional derivative of the effective action. It
contains the correlation lengths for superconductors or the Fermi surface or
correlated electrons in solids. This computation of the correlation length
requires a local coupling of the field to the source, which is usually assumed
tacitly.

In a quantum field theory context general relativity is based on the assumption
that the effective action for a macroscopic metric field takes the form of the
Einstein-Hilbert action with a cosmological constant. This is dictated by
diffeomorphism symmetry and an expansion in second order in derivatives. We
emphasize that this assumption concerns the effective action. The underlying
microscopic degrees of freedom of a theory of quantum gravity are not known. For
given microscopic degrees of freedom we do not know their precise relation to
the macroscopic metric. Also the classical action for some assumed microscopic
degrees of freedom of quantum gravity is not known. Nevertheless, a huge amount
of observables is derived from the assumed form of the effective action.

More generally, on the level of the effective action the direct contact to
microphysics may be lost. For example, the Kosterlitz-Thouless phase transition
occurs for many very different two-dimensional systems with continuous
$\text{SO}(2)$-symmetry. The universal critical behavior is encoded in an ansatz
for the effective action containing only a few functions defining the universal
equation of state. From this ansatz one can compute the macroscopic correlation
function and many other quantities. The precise relation to the microscopic
correlation functions for the different models realizing the Kosterlitz-Thouless
transition may be rather involved and is often not known in practice.
Nevertheless, the macroscopic correlation functions describe a good deal of the
relevant macroscopic observations.

Macroscopic field transformations are variable transformations of the
field-arguments of the effective action. For a given effective action they can
be performed without the knowledge of the underlying microscopic action. Such
macroscopic field transformations affect, however, the relation between
macroscopic and microscopic correlation functions. If one is interested in the
precise connection between macrophysics and microphysics one has to keep track
of this effect. This also concerns the issue of field transformations in flow
equations that we will discuss in the following sections.

In order to shed light on the precise relation between macrophysics and
microphysics we investigate in this section the computation of the effective
action from the classical action of a scalar quantum field theory. In this
context we can address the question what are the relation between macroscopic
and microscopic correlation functions and how they behave under field
transformations. For a choice of \qq{canonical fields} the relation between
macroscopic and microscopic correlation function follows directly from the
definition of the effective action. They are identical. Performing non-linear
microscopic or macroscopic field transformations this direct connection between
macroscopic and microscopic correlation functions will be lost. Often we do not
know the precise microscopic composite field whose expectation value yields the
macroscopic field. Nor do we know which precise choice of macroscopic field
corresponds to a canonical field. Our discussion highlights the complexity of
the precise relation between macrophysics and microphysics.

The issue of the macroscopic change of field variables for the effective action
$\Gamma$ differs conceptually from the microscopic field transformation on the
level of the functional integral discussed in the previous section. The
computation of $\Gamma$ performs already the functional integral. Once the
effective action is computed or its form assumed, no further integration over
fields is needed and the Jacobian for integration variables plays no longer a
role. The quantum field equation and the macroscopic correlation functions are
obtained by taking functional derivatives of the effective action. On this level
one can perform arbitrary transformations of field variables. They translate
directly to the functional derivatives of the effective action by the standard
rules for differentiation. By our assumption physical observables are expressed
as functions of the functional derivatives of the effective action or the
macroscopic correlation functions. The expectation values of observables remain
the same if the field transformations are inserted in the functional derivatives
of the effective action. Of course, the relation between observables and
macroscopic correlation functions change their functional form by this
insertion.

Field equations are differential equations. They can be solved for an arbitrary
choice of field-variables. Or course, for non-vanishing sources on has to take
the transformation of the source term into account. It transforms as
$\partial\Gamma/\partial\vp$. For a transformation from $\vp$ to $\vptil$,
defined by a function $\vp(\vptil)$, the transformed source $\tilde j$ for
$\vptil$ obeys $\tilde j = (\partial\vp/\partial\vptil ) j$, with $j$ the source
for $\vp$.

In short, the physics remains invariant under an arbitrary
change of macroscopic field-variables on which a given effective action depends.
We call the transformations of the variables in the effective action
\qq{macroscopic field transformations}. What is affected by the macroscopic
field transformations is the relation between the macroscopic correlation
functions and the microscopic correlation functions defined by expectation
values of powers of microscopic fields. We discuss this issue in detail.

\subsection*{Effective action as functional integral}

In place of the most commonly used definition of the effective action by the
Legendre transform of the logarithm of the partition function in the presence of
sources, we employ here an equivalent definition by an implicit functional
integral~\cite{Berges_2002} (background field integral),
\begin{align}
\label{E1}
&\Gamma[\chi]=\nn\\
&-\ln\int\cD\chi'\,\exp\left\{-S[\chi']+\int_x\frac{\partial\Gamma}{\partial\chi(x)}\gl\chi'(x)-\chi(x)\gr\right\}\
.
\end{align}
This is an integro-differential equation due to the presence of the sources
\bel{E2}
j(x)=\frac{\partial\Gamma}{\partial\chi(x)}\ .
\ee
The exact quantum field equation in the absence of sources is given by
\bel{E3}
\frac{\partial\Gamma}{\partial\chi(x)}=0\ .
\ee
In the form~\eqref{E1} it is apparent that a change of the integration variable
$\chi'(x)$ does not change the integral $\Gamma[\chi]$, provided that the
Jacobian is properly included.

Taking a functional derivative of eq.~\eqref{E1} with respect to $\chi(x)$
yields the identity
\bel{E4}
\int_y\Gamma^{(2)}(x,y)\gl\langle\chi'(y)\rangle-\chi(y)\gr=0\ .
\ee
Here we define the expectation value of an observable in the presence of sources
by
\bel{E5}
\langle
A\rangle=\frac{\int\cD\chi'\,A[\chi']\exp\big\{-S[\chi']+\int_xj(x)\gl\chi'(x)-\chi(x)\gr\big\}}{\int\cD\chi'\,\exp\big\{-S[\chi']+\int_xj(x)\gl\chi'(x)-\chi(x)\gr\big\}}\
,
\ee
and $\Gamma^{(2)}$ is the second functional derivative of $\Gamma$,
\bel{E6}
\Gamma^{(2)}(x,y)=\frac{\partial^2\Gamma}{\partial\chi(x)\partial\chi(y)}\ .
\ee
If $\Gamma^{(2)}$ is invertible, one concludes that the macroscopic field
$\chi(x)$ equals the expectation value of the microscopic field $\chi'(x)$ in
the presence of a suitable source
\bel{E7}
\chi(x)=\langle\chi'(x)\rangle\ .
\ee

Taking a derivative of eq.~\eqref{E7} with respect to $\chi(z)$ expresses the
connected two-point function $G(y,z)$ as the inverse of
$\Gamma^{(2)}$,
\bel{E8}
\int_y\Gamma^{(2)}(x,y)G(y,z)=\delta(x-z)\ ,
\ee
with $G$ given by the expectation value~\eqref{M5} in the presence of sources.
Taking further functional derivatives of eq.~\eqref{E8} the effective action is
found to be the generating functional for the one-particle-irreducible Green
functions for the field $\chi'(x)$.

In terms of the fluctuations around the macroscopic field,
\bel{E9}
\chat(x)=\chi'(x)-\chi(x)\ ,
\ee
we can write the effective action in the form
\bel{E10}
\Gamma[\chi]=-\ln\int\cD\chat\,\exp\left\{-S[\chi+\chat]+\int_x\frac{\partial\Gamma}{\partial\chi(x)}\chat(x)\right\}\
.
\ee
This step assumes that the measure $\cD\chat$ is defined such that
$\cD\chat=\cD\chi'$. In this form the saddle point approximation~\eqref{M21}
follows as an expansion of the argument of the exponential on the r.h.s. of
eq.~\eqref{E10} in powers of $\chat$. One inserts in lowest order
$\partial\Gamma/\partial\chi(x)=\partial S/\partial\chi(x)$, and performs the
Gaussian integral for the term quadratic in $\chat$. As before, the loop
integrals need a regularization.

A change of the integration variable from $\chi'$ to $\vp'$ does not change the
functional $\Gamma[\chi]$ defined by eq.~\eqref{E1}. Of course, for the
functional integral the Jacobian has to be taken into account. It becomes
apparent why in the preceding section the saddle point expansions for the
$\chi$-scheme and the $\vp$-scheme differ. In the $\vp$-scheme the step from
eq.~\eqref{E1} to eq.~\eqref{E10} requires to take out the Jacobian from the
measure such that $\int\cD\vp'=\prod_x\int\text{d}\vp(x)$ obeys
$\cD\hat\vp=\cD\vp'$. Once the Jacobian is included in the classical action,
$\Sbar[\vp']=S[\vp']-\ln J[\vp']$, and we express the source term
$\int_x\partial\Gamma/\partial\chi)(\chi'-\chi)$ in terms of $\vp'$, one obtains
the same $\Gamma$ for either choice of integration variable. The saddle point
expansions in the $\chi$- and $\vp$-schemes start, however, from a different
tree approximation $S[\chi']\neq\Sbar[\vp']$, and also have different loop
contributions since $S^{(2)}$ differs from $\Sbar^{(2)}$.

\subsection*{Transformations of microscopic and macroscopic fields}

For the \qq{canonical} field $\chi$ the macroscopic and the microscopic
correlation functions coincide. This is visible, for example, in the
relation~\eqref{E8} for which $(\Gamma^{(2)})^{-1}$ is the macroscopic connected
two-point function and $G$ the microscopic connected two-point function. For
non-linear field transformations this direct relation between macroscopic and
microscopic correlations is lost. In general, there is no longer any simple
relation between the macroscopic correlations for the transformed field and the
microscopic correlations for some field obtained by a microscopic field
transformation. The direct relation between macroscopic and microscopic
correlations is maintained only for linear field transformations.

As stated above, once the effective action is computed one is free to make any
macroscopic variable change from $\chi$ to $\vp[\chi]$. The derivative
$\partial\Gamma/\partial\vp(x)$ is easily expressed in terms of
$\partial\Gamma/\partial\chi(x)$, and similarly for higher functional
derivatives as $\partial^2\Gamma/\partial\vp(x)\partial\vp(y)$. The relation
between functional derivatives and correlation functions for microscopic fields
needs some care, however. Let us consider non-linear field transformations of
the macroscopic and microscopic fields
\bel{E11}
\vp(x)=N_{m}\gl\chi(x)\gr\chi(x)\ ,\quad \vp'(x)=N\gl\chi'(x)\gr\chi'(x)\ ,
\ee
with
\bel{E12}
\frac{\partial\vp}{\partial\chi}=\tilde N_{m}(\chi)\ ,\quad
\frac{\partial\vp'}{\partial\chi'}=\tilde N(\chi')\ ,\quad \tilde N_{m}
=N_{m}+\frac{\partial N_{m}}{\partial \ln\chi}\ .
\ee
Even though we may use the same function $N$ in both expressions in
eq.~\eqref{E11}, we stress that we deal with different transformations. For a
better distinction we use the index $m$ for the macroscopic field
transformations for which $N_m$ depends on the expectation value
$\chi=\langle\chi'\rangle$. In contrast, for the microscopic field
transformation $N$ is a function of the microscopic field $\chi'$.

We may compare the inverse of the second functional derivative of $\Gamma$ with
respect to $\vp$ and the connected two point function for $\vp'$. With
\begin{align}
\label{E13}
&\Gamma_{(\vp)}^{(2)}(x,y)=\frac{\partial^2\Gamma}{\partial\vp(x)\partial\vp(y)}\\
&=\tilde
N_{m}^{-1}(x)\frac{\partial^2\Gamma}{\partial\chi(x)\partial\chi(y)}\tilde
N_{m}^{-1}(y)+\frac{\partial^2\chi(y)}{\partial\vp(x)\partial\vp(y)}
\frac{\partial\Gamma}{\partial\chi(y)}\ ,\nn
\end{align}
one has for $\partial\Gamma/\partial\chi(y)=0$
\begin{align}
\label{E14}
\gl\Gamma_{(\vp)}^{(2)}\gr^{-1}(x,y)&=\tilde N_{m}(x)G(x,y)\tilde N_{m}(y)\\
&=\tilde N_{m}\gl\langle\chi'(x)\rangle\gr\tilde
N_{m}\gl\langle\chi'(y)\rangle\gr\nn\\
&\quad\times\gl\langle\chi'(x)\chi'(y)\rangle-
\langle\chi'(x)\rangle\langle\chi'(y)\rangle\gr\ .\nn
\end{align}
On the other side, the connected two-point function for $\vp'$ reads
\begin{align}
\label{E15}
G_{(\vp)}(x,y)=&\,\langle\vp'(x)\vp'(y)\rangle-\langle\vp'(x)\rangle\langle\vp'(y)\rangle\nn\\
=&\,\langle N\gl\chi'(x)\gr\chi'(x)N\gl\chi'(y)\gr\chi'(y)\rangle\nn\\
&-\langle N\gl\chi'(x)\gr\chi'(x)\rangle\langle N\gl\chi'(y)\gr\chi'(y)\rangle\
.
\end{align}
For non-linear field transformations this expression involves higher correlation
functions for $\chi'$. The expressions~\eqref{E14} and~\eqref{E15} differ even
if one uses the same functions for $N_m$ and $N$. Their explicit relation is not
obvious. A further difference arises for $\partial\Gamma/\partial\chi(y)\neq0$.
For non-linear field transformations with field dependent $N_m$ and $N$ the
relation between macroscopic and microscopic correlations is rather complex. The
relation~\eqref{E15} for the microscopic correlation function for $\vp'$
contains higher correlations for $\chi'$. They can be expressed in terms of
functional derivatives of $\Gamma$ with respect to the canonical field $\chi$.
In turn, these macroscopic correlations for $\chi$ can be expressed in terms of
macroscopic correlations for $\vp$. In the other direction, the macroscopic
correlation function $(\Gamma_{(\vp)}^{(2)})^{-1}$ involves the microscopic
two-point function for $\chi'$ and functions of expectation values
$\langle\chi'\rangle = \chi$. We can express $\chi$ in terms of $\vp$ by the
field transformation. The complication of a relation to microscopic correlations
for $\vp'$ arises for the complicated relation between correlations for $\vp'$
and $\chi'$, as in eq.~\eqref{E15}. In summary, if one wants to keep precise
track of microscopic correlations the use of non-linear field transformations is
perhaps not a very good idea.

In summary, we have defined the effective action $\Gamma_\chi[\chi]$ by the
implicit relation~\eqref{E1}. Here the index $\chi$ of $\Gamma$ indicates the
form of the source term linear in $\chi'-\chi$. The correlation functions for
the microscopic field $\chi'$ are simply related to functional derivatives of
$\Gamma_\chi[\chi]$ with respect to $\chi$. By a macroscopic field
transformation $\chi=\chi[\vp]$ we obtain
\bel{E15A}
\Gamma_\chi[\vp]=\Gamma_\chi\big[\chi[\vp]\big]\ .
\ee
The functional derivatives of $\Gamma_\chi[\vp]$ with respect to $\vp$ do not
generate the correlation functions for the microscopic field $\vp'$ which
obtains from $\chi'$ by an analogous field transformation. They rather multiply
the correlation functions for $\chi'$ with suitable functions of the expectation
value $\chi$. In this sense $\chi$ is singled out as the \qq{canonical field}
for
the effective action $\Gamma_\chi$. Fields $\vp[\chi]$ obtained by a non-linear
field transformation are no longer canonical as far as the relation of
functional derivatives of $\Gamma_\chi[\vp]$ to correlation functions of
microscopic fields are concerned.

\subsection*{Macroscopic correlation functions}

We define \qq{macroscopic correlation functions} for $\vp$ by suitable
functional derivatives of $\Gamma_\chi$ with respect to $\vp(x)$. For example,
the macroscopic connected two-point function for $\vp$ is given by the inverse
of $\Gamma_{(\vp)}^{(2)}$ in eq.~\eqref{E13}. In contrast, the \qq{microscopic
correlation functions} are given by suitable expectation values of powers of
the microscopic field $\vp'(x)$. For the canonical field $\chi'$ the macroscopic
and microscopic correlation functions coincide. For the non-canonical field
$\vp$ or $\vp'$ we have seen that in general the macroscopic and microscopic
correlation functions differ even if we choose the same function $N_m$ and $N$.

As we have argued at the beginning of this section for many problems the
knowledge of the microscopic correlation function plays actually no role. Only
the macroscopic correlation functions matter, and therefore only suitable
macroscopic variable transformations in the effective action $\Gamma_\chi$ are
involved. As an example, we may consider the field equation in the presence of a
source,
\bel{E25}
\frac{\partial\Gamma_\chi}{\partial\chi(x)}=j(x)\ .
\ee
This translates directly to the field equation for $\vp(x)$
\bel{E26}
\frac{\partial\Gamma_\chi}{\partial\vp(x)}=\tilde N_{m}^{-1}\gl\vp(x)\gr
j(x)=j_\vp(x)\ .
\ee
Let us denote the solution of the field equation in the absence of sources by
$\vp_0(x)=N_{m}\gl\chi_0(x)\gr\chi_0(x)$, where
$\partial\Gamma/\partial\chi(x)(\chi_0)=0$. For small $j_\vp(x)$ one can solve
the linearized field equation, with solution denoted by
$\bar\vp(x)=\vp_0(x)+\delta\vp(x)$,
\begin{align}
\label{E27}
\int_y\frac{\partial^2\Gamma}{\partial\vp(x)\partial\vp(y)}&\delta\vp(y)=j_\vp(x)\
,\nn\\
\delta\vp(x)=&\,\int_y\gl\Gamma_\vp^{(2)}\gr^{-1}(x,y)j_\vp(y)\ .
\end{align}
The linearized solution only involves the macroscopic correlation function
$(\Gamma_\vp^{(2)})^{-1}$. It is the macroscopic rather than the microscopic
correlation function which describes the response of the macroscopic system, as
characterized by the value of $\vp$, to a small (macroscopic) source.

The role of the macroscopic correlation functions is particularly apparent for
critical systems. Many, sometimes vastly different, microscopic models can lead
to the same universal critical behavior which is encoded in the form of the free
energy of effective action $\Gamma$. Important quantities as the correlation
length can be extracted directly from $\Gamma$ for an many choices of fields,
reflecting properties of the second functional derivative $\Gamma^{(2)}$. This
only needs locality of the coupling of fields to sources. The latter is
maintained by local non-linear field transformations. The microscopic
correlation functions play no role here -- they can be actually very different
for the different microscopic models. Only the macroscopic fields play a role
for the universal critical behavior. In terms of the microscopic fields of a
given microscopic model they are often rather complex composite fields.

In summary, on the level of the effective action a detailed understanding of
field transformations of microscopic variables is only needed for the relation
between macroscopic and microscopic correlation functions. As long as only the
macroscopic level is concerned we can simply consider non-linear transformations
of the macroscopic fields as variable transformations for a given functional and
its derivatives. It is the analogue to variable transformations for arguments of
a function.

\subsection*{Different effective actions}

The choice of the effective action is not unique. Different effective actions
can be defined by different source terms. For example, we may specify a
different effective action $\Gamma_\vp[\vp]$ by replacing in eq.~\eqref{E1} the
source term by $\int_x\partial\Gamma/\partial\vp(x)\gl\vp'(x)-\vp(x)\gr$.
This corresponds to a canonical source term for the \qq{composite field} $\vp'$,
\begin{align}
\label{E27A}
&\Gamma_\vp[\vp]=\nn\\
&-\ln\int\cD\chi'\,\exp\left\{-S[\chi']+\int_x\frac{\partial\Gamma_\vp}{\partial\vp(x)}\gl\vp'(x)-\vp(x)\gr\right\}\
.
\end{align}
The macroscopic correlation functions for $\vp$ derived from $\Gamma_\vp[\vp]$
equal the microscopic correlation functions for $\vp'$. The difference to the
previous version of the effective action $\Gamma_\chi[\vp]$ arises only from the
different source term. (Both $\Gamma_\chi[\vp]$ and $\Gamma_\vp[\vp]$ are
independent of the choice of the integration variable.) As far as the relation
between macroscopic and microscopic correlation functions is concerned the
effective action admits a \qq{canonical field choice}. The canonical field is
$\chi(x)$ for $\Gamma_\chi$ and $\vp(x)$ for $\Gamma_\vp$. Only for the
canonical field choice the macroscopic correlation functions equal the
microscopic ones.

\subsection*{Effective action for fundamental and composite fields}

Further light on this issue can be shed by introducing sources for both $\chi'$
and $\vp'$ and defining an effective action depending both on the macroscopic
fields $\chi$ and $\vp$, treated as independent variables,
\begin{align}
\label{E28}
&\Gamma[\chi,\vp]=\nn\\
&-\ln\int\cD\chi'\,\exp\bigg\{-S[\chi']+\int_x\bigg[\frac{\partial\Gamma}{\partial\chi(x)}\gl\chi'(x)-\chi(x)\gr\nn\\
&\quad+\frac{\partial\Gamma}{\partial\vp(x)}\gl
N\gl\chi'(x)\gr\chi'(x)-\vp(x)\gr\bigg]\bigg\}\ .
\end{align}
This results in
\bel{E29}
\chi(x)=\langle\chi'(x)\rangle\ ,\quad \vp(x)=\langle\vp'(x)\rangle=\langle
N\gl\chi'(x)\gr\chi'(x)\rangle\ ,
\ee
where expectation values are taken in the presence of the two sources
\bel{E30}
\tilde j_\chi(x)=\frac{\partial\Gamma}{\partial\chi(x)}\ ,\quad \tilde
j_\vp(x)=\frac{\partial\Gamma}{\partial\vp(x)}\ .
\ee

Let us denote by $\vp_0[\chi]$ the solution of the partial field equation
\bel{E31}
\frac{\partial\Gamma[\chi,\vp]}{\partial\vp(x)}\gl\vp_0[\chi]\gr=0\ .
\ee
This defines a macroscopic field transformation by
\bel{66A}
\vp_0[\chi]=N_{m}(\chi)\chi\ .
\ee
In general, the function $N_{m}$ differs from $N$ as used in the microscopic
field transformation in eqs.~\eqref{E28}~\eqref{E29}.

Inserting this $\chi$-dependent solution into eq.~\eqref{E28} we recover
\bel{E32}
\Gamma_\chi[\chi]=\Gamma\big[\chi,\vp_0[\chi]\big]\ .
\ee
Defining similarly $\chi_0[\vp]$ by the solution of the partial field equation
\bel{E33}
\frac{\partial\Gamma[\chi,\vp]}{\partial\chi(x)}\gl\chi_0[\vp]\gr=0\ ,
\ee
yields $\Gamma_\vp[\vp]$. In general, $\Gamma_\vp[\vp]$ and $\Gamma_\chi[\vp]$
differ,
\begin{align}
\label{E35}
\Gamma_\chi[\vp]&=\Gamma\left[\frac{\vp}{N_{m}(\vp)},\vp_0\bigg[
\frac{\vp}{N_{m}(\vp)}\bigg]\right]\ ,\nn\\
\Gamma_\vp[\vp]&=\Gamma\big[\chi_0[\vp],\vp\big]\ .
\end{align}
Nevertheless, the two functionals $\Gamma_\chi$ and $\Gamma_\vp$ coincide on the
solution $(\chi_0,\vp_0)$ of the full field equation
\bel{E36}
\frac{\partial\Gamma}{\partial\chi}[\chi_0,\vp_0]=0\ ,\quad
\frac{\partial\Gamma}{\partial\vp}[\chi_0,\vp_0]=0\ ,
\ee
where
\bel{E37}
\Gamma_\chi[\vp_0]=\Gamma_\vp[\vp_0]=\Gamma[\chi_0,\vp_0]\ .
\ee

Besides the conceptual clarification an effective action depending both on a
field $\chi$ and a \qq{composite} of this field $\vp[\chi]$ is often useful for
practical purposes. An example are strong interactions in elementary particle
physics where meson fields can be introduced as bilinears in the quark fields.

\subsection*{Frame invariance}

For a given effective action $\Gamma_\chi$ or $\Gamma_\vp$ the fields $\chi'(x)$
or $\vp'(x)$ can be considered as different coordinates for field-space, and
similarly for $\chi(x)$ or $\vp(x)$. Observables are independent of the
coordinate choice or choice of \qq{frame}. This frame independence can be
formulated on different levels. Microscopic frame invariance states that
observables are independent of the choice of microscopic fields $\chi'(x)$ or
$\vp'(x)$. A given choice of effective action as $\Gamma_\chi[\chi]$ or
$\Gamma_\vp[\vp]$ is per se invariant under microscopic frame transformations
since those only concern the change of the integration variables in an integral.
This supposes, of course, that the Jacobian is properly taken into account. All
correlation functions and observables that can be computed from the effective
action take the same values in all microscopic frames. For this type of
microscopic frame transformations the sources are kept fixed, i.e.
$j(x)=\partial\Gamma_\chi/\partial\chi(x)$ in the functional integral~\eqref{E1}
is not changed. As a consequence the source term changes its form if $\chi'$ is
replaced by $\chi'(\vp')$. A different form of microscopic frame invariance, for
which the sources are transformed simultaneously, will be discussed below.

Macroscopic frame transformations concern field transformations of the
macroscopic fields $\chi(x)$ or $\vp(x)$. Observables that can be expressed by
macroscopic correlation functions can equally be computed from
$\Gamma_\chi[\chi]$ or $\Gamma_\chi[\vp]=\Gamma_\chi\big[\chi[\vp]\big]$. They
take the same values provided we replace properly in all functional derivatives
\bel{E38}
\frac{\partial}{\partial\chi(x)}=\int_y\frac{\partial\vp(y)}{\partial\chi(x)}\frac{\partial}{\partial\vp(y)}\
.
\ee
This simple statement is macroscopic frame invariance. For invertible field
transformations it implies directly that the quantum field equations in the
absence of sources can be equally computed from $\Gamma_\chi[\chi]$ or
$\Gamma_\chi[\vp]$,
\bel{E39}
\frac{\partial\Gamma_\chi[\chi]}{\partial\chi(x)}=
0\iff\frac{\partial\Gamma_\chi[\vp]}{\partial\vp(x)}=0\ .
\ee
Macroscopic frame invariance extends to field equations in the presence of
sources if the sources are transformed properly according to eq.~\eqref{E26}.

Macroscopic frame invariance is relevant for cosmology. In any theory of quantum
gravity the relevant field equations are the quantum field equations obtained by
variation of the quantum effective action. The relevant fields are the
macroscopic metric together with possible additional macroscopic fields as
scalar fields. Different \qq{metric frames} correspond to different choices of
the macroscopic metric field. They are related by macroscopic frame
transformations. An example are the well known Weyl scalings~\cite{HWGE,
PhysRev.125.2163} or conformal transformations which multiply the metric by a
function of a scalar field. Macroscopic frame invariance states that the quantum
field equations derived from variation of the effective action are equivalent
for all metric frames. This equivalence of cosmological equations on the quantum
level has been advocated for a long time~\cite{CWVNC}.

In contrast, in a quantum context a similar equivalence does not hold for the
classical field equations. Two classical actions $S[\chi']$ and
$S[\vp']=S\big[\chi'[\vp']\big]$ lead, of course, to the same classical field
equations. As a starting point for a quantum theory the two actions are not
equivalent, however, unless supplemented by an appropriate Jacobian in the
functional measure. This Jacobian leads effectively to different classical
actions $S[\chi']$ and $\Sbar[\vp']$. Even if the Jacobian is included (which is
usually not done) the loop expansion in any finite order will lead to different
results, as we have seen in section~\ref{sec:FTFI}. In the past, these facts
have led to some confusion about the equivalence of different metric frames.

Beyond the quantum field equations, the macroscopic frame invariance also
extends to the characterization of fluctuations. The \qq{primordial cosmic
fluctuations} in inflationary cosmology correspond to the macroscopic
correlation function, with power spectrum directly proportional to the
macroscopic connected two-point function in momentum space. It is rather obvious
that the metric fluctuations in the observable range of scales correspond to
fluctuations of some type of macroscopic metric. The details of microscopic
correlation functions for degrees of freedom defined at length scales much
shorter than the Planck length play no role - the macroscopic metric could even
be a composite of different microscopic fields, as scalar or vector fields.
Details of the relation between microscopic and macroscopic correlations could
only play a role if they affect the form of the effective action.

The primordial cosmic fluctuation spectrum can indeed be computed by inverting
the second functional derivative of the effective action~\cite{CWCF, CWQCM}. The
power spectrum of primordial fluctuations can be shown directly to be a frame
invariant quantity with respect to Weyl scalings~\cite{Wetterich_2016}. In the
context of quantum gravity and cosmology the macroscopic frame invariance has
been named \qq{field relativity}~\cite{CWUE, Wetterich_2015}. Since geometry
depends on the choice of the metric, macroscopic frame invariance implies the
relativity of geometry. Geometry remains invariant only under the particular
metric transformations that correspond to a change of the coordinates of
space-time (diffeomorphisms). Under more general metric transformations, as Weyl
scalings, the geometry changes.

\subsection*{(In)equivalent source terms}

A more subtle issue concerns field transformations accompanied by changes of
source terms in the definition of the effective action. They are also often
named frame transformations. For this version of microscopic frame
transformations the effective action changes due to the change of the source
term. While a macroscopic frame transformation changes
$\Gamma_\chi[\chi]\to\Gamma_\chi[\vp]$, this version of a microscopic frame
transformation amounts to the map $\Gamma_\chi[\chi]\to\Gamma_\vp[\vp]$. It is
this version of microscopic frame transformations that will be generalized to
the microscopic field transformations of flow equations~\cite{PAW} which are the
basis of the \qq{essential renormalization group}~\cite{BZF, BAFA}.

This type of microscopic frame transformation maps the functional
integral~\eqref{E1} for $\Gamma_\chi[\chi]$ to the functional
integral~\eqref{E27A} for $\Gamma_\vp[\vp]$ or
\begin{align}
\label{73A}
\Gamma_\vp[\vp]=&\,-\ln\int\cD\vp'\,\exp\bigg\{-\Sbar[\vp']\nn\\
&+\int_x\frac{\partial\Gamma_\vp}{\partial\vp(x)}\gl\vp'(x)-\vp(x)\gr\bigg\}\ .
\end{align}
Both microscopic and macroscopic fields are transformed. A change of the
integration variable has no effect on the effective action. The change of the
macroscopic field is a change of the variables in terms of which the effective
action is expressed. The core of this type of microscopic frame transformation
is therefore the non-trivial transformation of the source term.

 A possible view considers the source terms only as an auxiliary construction
and evaluates correlation functions at vanishing sources $j(x)=0$, or
equivalently for solutions of the field equations obtained by variation of the
effective action. The final result for microscopic correlation functions is then
evaluated by a functional integral in the absence of a source term according to
eq.~\eqref{E5} with $j(x)=0$. These correlation functions are, of course,
independent of the precise definition of the source term, provided that the
final result is obtained for a vanishing source term.

On the other hand, we have seen explicitly that the effective actions
$\Gamma_\chi[\vp]$ and $\Gamma_\vp[\vp]$, which are distinguished by different
source terms, are different functionals. They describe different relations
between microscopic and macroscopic correlation functions. In principle, the
microscopic correlation functions for a given microscopic model defined by the
classical action $S[\chi']$ and the functional measure $\cD\chi'$ can be
formally inferred both from $\Gamma_\chi[\vp]$ and from $\Gamma_\vp[\vp]$. In
practice, the way how they are derived can be rather different, however. The
difference of the functionals $\Gamma_\chi[\vp]$ and $\Gamma_\vp[\vp]$ has to be
compensated by different relations how a given observable or microscopic
correlation function in the absence of sources is related to functional
derivatives of $\Gamma_\chi[\vp]$ or $\Gamma_\vp[\vp]$. While this relation
exists formally, it is typically much too complicated to be of practical use.
Whenever one wants to exploit the behavior of the effective action away from the
solution of the quantum field equations one has to consider $\Gamma_\chi[\vp]$
and $\Gamma_\vp[\vp]$ as different functionals. Only the field equations in the
absence of sources, and the values of the effective action for solutions of the
field equations are identical for $\Gamma_\chi[\vp]$ and $\Gamma_\vp[\vp]$.

These remarks are relevant for the macroscopic correlation functions evaluated
on solutions of the field equations. The second functional derivatives of
$\Gamma_\chi[\vp]$ and $\Gamma_\vp[\vp]$ differ, as we have demonstrated
explicitly. The question arises which effective action should be used for the
definition of macroscopic correlation functions. In principle, this is
determined by the question which macroscopic correlation functions are relevant
for a given observation.

For many questions the difference between $\Gamma_\chi[\vp]$ and
$\Gamma_\vp[\vp]$ plays no role for real observations in practice. This is
related to the fact that observations typically concern length scales much
larger than the microscopic length scales on which the theory is defined.
Universality induced by the renormalization flow washes out differences in the
microscopic definition. In the momentum range relevant for observations
$\Gamma_\chi[\vp]$ and $\Gamma_\vp[\vp]$ are typically the same, or
distinguished only by somewhat different values of renormalizable couplings. One
can then use $\Gamma_\chi[\vp]$ or $\Gamma_\vp[\vp]$ equivalently for the
computation of macroscopic correlation functions.

The main conclusions of this section state that macroscopic field
transformations (frame transformations) for the arguments of the quantum
effective action do not affect the physics. Macroscopic correlation functions
obtain by functional derivatives of the effective action and transform
covariantly. They are relevant for many situations of interest. Only if one
wants to establish a relation between the macroscopic correlation functions and
the microscopic correlation functions a canonical choice of fields (canonical
frame) is singled out.

Different versions of the quantum effective action
correspond to different choices of source terms. They are related by a type of
microscopic frame transformation which also transforms the source term. These
different versions are strictly equivalent as far as solutions of the quantum
field equations in the absence of sources are concerned. They often remain
(approximately) equivalent for the macroscopic correlation functions as well.
Nevertheless, on the conceptional level they differ. The microscopic field
transformations change the effective action, while the fields remain canonical.
The macroscopic field transformations leave the effective action invariant,
while a canonical field does not remain canonical.

\section{Field transformations in flow equation}\label{sec:FTFE}

A conceptually, and often also practically, convenient way to compute the
effective action employs a functional flow equation. In this approach the
fluctuation effects are included stepwise by variation of a suitable cutoff. The
change of the effective action for an infinitesimal step typically involves a
momentum integration which covers effectively only a finite momentum range. As a
consequence, the loop integrals are finite and no further regularization is
needed. The cutoff involves a characteristic momentum scale $k$. The change of
the effective action by a variation of $k$ obeys an exact functional flow
equation~\cite{Wetterich_1993}. Conceptually, one starts at very high $k$ (or
infinite $k$) with the microscopic effective action. This should be related to
the \qq{classical action} which defines the functional integral. Following the
flow equations to $k\to0$ yields the quantum effective action which includes all
fluctuation effects. In practice, a truncation of the form of the effective
action is needed in order to obtain approximate solutions of the functional flow
equation.

Field transformations for the flow equation can be implemented in several ways.
\qq{Macroscopic field transformations}~\cite{CWFT} amount to variable
transformations for a given flow equation. They correspond to a simple change of
variables for a differential equation. In contrast, \qq{microscopic field
transformations}~\cite{PAW} change the cutoff and the source term used for the
definition of the flow equation. A microscopic field transformation therefore
corresponds to a map onto a different flow equation which is not related to the
original flow equation by a simple change of variables. Microscopic variable
transformations also affect the relation between the classical action and the
microscopic effective action. This brings back indirectly the issue of the
Jacobian in the functional integral. Despite their different conceptual status
the macroscopic and microscopic variable transformations have in practice often
very similar effects. They are equivalent for linear field transformations. We
focus in this section on the macroscopic field transformations and discuss the
microscopic field transformations in sects.~\ref{sec:FECS},~\ref{sec:TFF}.

We demonstrate many of the issues encountered for field transformations by a
particular example. This concerns the flow of the field-dependent coefficient of
the derivative term (kinetial) in a simple scalar field theory. For comparison,
we present in two appendices~\ref{app:FEK} and~\ref{app:FKP} the flow equation
for the kinetial without performing a field transformation. In
appendix~\ref{app:FEK} we omit the effect of the scalar potential on the flow of
the kinetial. The contributions of the potential are included in
appendix~\ref{app:FKP}. These computations can serve as a benchmark for
comparison with the results obtained by field transformations in the flow
equation. We will see that the flow equations are not changed by $k$-independent
field transformations. The computations of the
appendices~\ref{app:FEK},~\ref{app:FKP} therefore cover this case as well. We
find strong indications for the existence of the ultraviolet fixed
points~\eqref{eq:AA} from these computations.

\subsection*{Effective average action and flow equation}

We first present an intuitive definition of the effective average action
$\Gamma_k$ in complete analogy to eq.~\eqref{E10}, adding only a cutoff term for
the fluctuations
\begin{align}
\label{F1}
\Gamma_{\chi,k}[\chi]=-\ln\int\cD\chat\,\exp\bigg\{&-S[\chi+\chat]+\int_x\frac{\partial\Gamma_k}{\partial\chi(x)}\chat(x)\nn\\
-\frac12\int_{x,y}&\chat(x)\cR_k(x,y)\chat(y)\bigg\}\ .
\end{align}
The term $\sim\cR_k$ serves as a cutoff, typically an infrared cutoff which
suppresses fluctuations with $q^2<k^2$. It introduces the renormalization scale
$k$. We assume $\cR_{k=0}=0$, such that for $k=0$ the effective average action
becomes the quantum effective action $\Gamma_\chi[\chi]$ defined by
eq.~\eqref{E10}.

While intuitive, the relation~\eqref{F1} is not the basic definition of the
effective average action. The basic definition of $\Gamma_k$ is similar to
eq.~\eqref{E1},
\begin{align}
\label{F2}
\Gamma_k[\chi]=&\,\Gamma_k'[\chi]-\Delta_kS[\chi]\ ,\nn\\
\Gamma_k'[\chi]=&\,-\ln\int\cD\chi'\,\exp\bigg\{-S[\chi']-\Delta_kS[\chi']\nn\\
&\quad\quad+\int_x\frac{\partial\Gamma_k'}{\partial\chi(x)}\gl\chi'(x)-\chi(x)\gr\bigg\}\
,
\end{align}
where the cutoff term $\Delta_kS$ is quadratic in $\chi'$
\bel{F3}
\Delta_kS[\chi']=\frac12\int_{x,y}\chi'(x)\cR_k(x,y)\chi'(y)\ .
\ee
Here $\Gamma_k'$ is equivalent to the Legendre transform of the Schwinger
functional for a microscopic action including the cutoff term. Subtracting this
cutoff term is important for making contact to the microscopic action and to
implement the simple one-loop form of the exact flow equation.

The definitions~\eqref{F1} and~\eqref{F2} are identical only for
$\cD\chat=\cD\chi'$. In this case they are related by a simple shift of
integration variables in the exponent, employing with $\cR_k(x,y)=\cR_k(y,x)$
the
relation
\bel{F4}
\frac{\partial\Gamma_k'}{\partial\chi(x)}=\frac{\partial\Gamma_k}{\partial\chi(x)}+\int_y\cR(x,y)\chi(y)\
.
\ee
For non-linear functional measures $\cD\hat\chi$ differs from $\cD\chi$ and the
relation~\eqref{F1} is no longer valid.

The relation~\eqref{F1} is a central tool for establishing that the microscopic
effective action $\Gamma_{k\to\infty}$ equals the classical action. This follows
directly if $\cR_k$ diverges for $k\to\infty$. The linearity of the functional
measure does no longer hold after a non-linear microscopic field transformation.
As a consequence, the microscopic effective action and the classical action are
no longer identical. We will see in sect.~\ref{sec:TFF} that their relation
brings back the Jacobian even if we do not change the functional measure. It is
the linearity of the functional measure, together with the linearity of the
source term, that singles out a canonical choice of fields for the flow
equation.

The definition~\eqref{F2} continues to hold even if $\cD\chat=\cD\chi'$ is not
realized. For $k=0$ one recovers
$\Gamma_k[\chi]=\Gamma_k'[\chi]=\Gamma_\chi[\chi]$, as defined by
eq.~\eqref{E1}. Observing that $\Gamma_k'$ is the effective action for a
modified classical action $S+\Delta_kS$ we obtain the relations
\bel{F5}
\langle\chi'(x)\rangle=\chi(x)\ ,\quad \langle\chat(x)\rangle=0\ ,
\ee
and
\begin{align}
\label{F6}
G(x,y)=&\,\langle\chat(x)\chat(y)\rangle=\gl\Gamma_k^{\prime(2)}\gr^{-1}(x,y)\nn\\
=&\,\gl\Gamma_k^{(2)}+\cR_k\gr^{-1}(x,y)\ .
\end{align}
Here the expectation values are evaluated for the classical action $S+\Delta_kS$
in the presence of the source $\partial\Gamma'/\partial\chi(x)$. For canonical
fields one can use equivalently the weight factor given by the exponential in
eq.~\eqref{F1}.

An exact functional flow equation is obtained by taking a $k$-derivative of
eq.~\eqref{F1},
\bel{F7}
\partial_k\Gamma_k=\frac12\int_{x,y}\partial_k\cR_k(x,y)\langle\chat(y)\chat(x)\rangle-\partial_k\int_x\frac{\partial\Gamma_k}{\partial\chi(x)}\langle\chat(x)\rangle\
.
\ee
Inserting the relations~\eqref{F5},~\eqref{F6} one obtains
\bel{F8}
\partial_k\Gamma_k=\frac12\tr\left\{\partial_k\cR_k\gl\Gamma_k^{(2)}+\cR_k\gr^{-1}\right\}\
.
\ee
Expressing the trace in momentum space and choosing diagonal $\cR_k$,
\begin{align}
\label{F9}
\cR_k(q,q')=&\,R_k(q^2)\delta(q,q')\ ,\nn\\
\delta(q,q')=&\,(2\pi)^d\prod_\mu\delta(q_\mu-q'_\mu)\ ,
\end{align}
yields
\bel{F10}
\partial_k\Gamma_k=\frac12\int_q\partial_kR_k(q^2)\gl\Gamma_k^{(2)}+\cR_k\gr^{-1}(q,q)\
,
\ee
where
\bel{F11}
\int_q=\frac1{(2\pi)^d}\prod_\mu\int_{-\infty}^\infty\text{d}q_\mu\ .
\ee
The single momentum integral reflects the one-loop form of the exact flow
equation. The effects of higher loops arise from the non-trivial field- and
momentum dependence of $\Gamma_k^{(2)}$.

For the example of diagonal $\Gamma_k^{(2)}$,
\bel{F12}
\Gamma_k^{(2)}(q,q')=\gl q^2+m^2\gr\delta(q,q')\ ,
\ee
the flow equation becomes
\bel{F13}
\partial_k\Gamma_k=\frac12\int_x\int_q\partial_kR_k(q^2)\gl
q^2+m^2+R_k(q^2)\gr^{-1}\ ,
\ee
where we employ with total volume of space $\Omega$
\bel{F14}
\delta(q,q)=\Omega=\int_x\ .
\ee
The diagonal form~\eqref{F12} typically obtains for $x$-independent macroscopic
fields $\chi(x)=\chi$, with $m^2$ depending on $\chi$. Constant $\chi$ yields
the part of $\Gamma_k$ which corresponds to the effective potential,
$\Gamma_k=\int_xU_k(\chi)+\hdots$. The momentum integral~\eqref{F13} is
ultraviolet finite if $R_k$ decays sufficiently fast for $q^2\to\infty$, and
infrared finite for a suitable infrared cutoff $R_k(q^2)$.

The finiteness of the flow equation generalizes to arbitrary $x$-dependent
macroscopic fields $\chi(x)$. The r.h.s. of the flow equation~\eqref{F8} is well
defined and needs no regularization. In this sense the flow equation constitutes
its own \qq{ERGE-regularization scheme}. The regularization conditions in the
loop expansion are replaced by the initial value of $\Gamma_k$ for very large
$k$, e.g. $k\to\infty$ or $k=\Lambda$, with $\Lambda$ some intrinsic UV-cutoff
scale as for lattice models $\Lambda\sim\eps^{-1}$.

\subsection*{Macroscopic field transformations}

The flow equation is a functional differential equation. As such, we can solve
it for an arbitrary choice of field variables. Often it is useful to employ
$k$-dependent field variables
\bel{F15}
\vp=\vp_k[\chi]\ .
\ee
This corresponds to a macroscopic $k$-dependent field transformation in the
effective average action. Evaluating the flow equation at fixed $\vp$ instead
of fixed $\chi$ induces a modification of the flow equation~\cite{CWFT},
\bel{F16}
\partial_t\Gamma_k|_\vp=\partial_t\Gamma_k|_\chi-\int_x\frac{\partial\Gamma_k}{\partial\vp(x)}\partial_t\vp(x)|_\chi\
.
\ee
Here we use the flow variable $t=\ln(k/k_0)$, $\partial_t=k\partial_k$, and
\bel{F17}
\partial_t\vp(x)|_\chi=k\frac{\partial\vp_k[\chi]}{\partial
k}(x)=\tilde\gamma_k[\vp]\ .
\ee
The last identity expresses $\partial_t\vp|_\chi$ as a functional of $\vp$. The
original flow equation~\eqref{F8} obtains for canonical fields. For composite
fields depending on $k$ an additional term appears in eq.~\eqref{F16}.

A simple example concerns the dimensionless \qq{scale invariant scalar fields}
in
four dimensions,
\bel{F18}
\vp(x)=\tilde\chi(x)=\frac{\chi(x)}{k}\ ,\quad \partial_t\vp(x)|_\chi=-\vp(x)\ ,
\ee
such that a \qq{scaling form} of the flow equation reads
\begin{align}
\label{F19}
\bigg(\partial_t-\int_x\tilde\chi(x)&\frac{\partial}{\partial\tilde\chi(x)}\bigg)\Gamma_k\nn\\
&=\frac12\tr\left\{\gl2+\partial_t\gr\tilde r_k\gl\tilde\Gamma_k^{(2)}+\tilde
r_k\gr^{-1}\right\}\ ,
\end{align}
where $\Gamma_k=\Gamma_k[\tilde\chi]$ and
\bel{F20}
\Gamma_k^{(2)}=k^2\tilde\Gamma_k^{(2)}\ ,\quad \cR_k=k^2\tilde r_k\ .
\ee
In close analogy we can define renormalized scalar fields by factoring out a
$k$-dependent wave function renormalization.

A second example involves momentum dependent rescalings of the Fourier modes
$\chi(q)$ in momentum space
\bel{F21}
\vp(q)=N_{m}\left(\frac{q^2}{k^2}\right)\chi(q)=N_{m}(y)\chi(q)\ ,\quad
y=\frac{q^2}{k^2}\ .
\ee
With
\bel{F22}
\partial_t\vp(q)|_\chi=-2\frac{\partial\ln N_{m}}{\partial\ln y}\vp(q)\ ,
\ee
the flow equation for $\vp$ reads
\begin{align}
\label{F23}
&\left(\partial_t+2\int_q\frac{\partial\ln N_{m}}{\partial\ln
y}(q)\vp(q)\frac{\partial}{\partial\vp(q)}\right)\Gamma_k[\vp]\nn\\
&=\frac12\int_q\left(\partial_t-4\frac{\partial\ln N_{m}}{\partial\ln
y}\right)\tilde R_k(q)\gl\tilde\Gamma_k^{(2)}+\tilde \cR_k\gr^{-1}(q,q)\ ,
\end{align}
where now
\begin{align}
\label{F24}
\tilde\Gamma_k^{(2)}(q,q')=&\,
\frac{\partial^2\Gamma_k}{\partial\vp(q)\partial\vp(q')}
=N_{m}^{-1}(q)N_{m}^{-1}(q')\Gamma_k^{(2)}(q,q')\ ,\nn\\ \tilde
\cR_k(q,q')=&\,N_{m}^{-1}(q)N_{m}^{-1}(q')R_k(q^2)\delta(q,q')\ .
\end{align}
Suitable forms of $N_{m}(q)$ may be used in order to absorb some of the momentum
dependence of $\Gamma_k^{(2)}(q,q')$ into a simple $\tilde\Gamma_k^{(2)}(q,q')$.
We emphasize, however, that vertices corresponding to derivatives with respect
to $\vp(q)$ receive now additional factors of $N_{m}(q)$. The momentum
dependence of
the propagator is shifted to the vertices.

For linear macroscopic field transformations we can adapt the cutoff function
such that it becomes simple in terms of the composite fields. For our example
$\tilde\cR_k$ can be chosen to be a simple function of $q^2$. A linear field
transformation does not change the quadratic form of the cutoff term in the
functional integral and does not induce a field-dependent Jacobian. We will see
in the next section that eq.~\eqref{F23} obtains equivalently from a microscopic
field transformation.

\subsection*{Non-linear field transformations}

For a non-linear macroscopic field transformation the modified flow equation
shows new more complex features. This is due to the fact that
$\partial_t\Gamma_k|_\chi$ has to be expressed in terms of $\vp$. Similarly, the
quantity $\tilde\gamma_k[\vp]$ may not always remain simple. In particular, the
flow equation involves now a connection term. For
\bel{F24A}
\frac{\partial}{\partial\chi(x)}=\int_u\tilde
N_{m}(x,u)\frac{\partial}{\partial\vp(u)}\ ,\quad \tilde
N_{m}(x,u)=\frac{\partial\vp(u)}{\partial\chi(x)}\ ,
\ee
the second derivative of $\Gamma$ with respect to $\chi$ results in a type of
covariant derivative with respect to $\vp$,
\begin{align}
\label{F25}
&\frac{\partial^2\Gamma_k}{\partial\chi(x)\partial\chi(y)}=\int_{u,v}\tilde
N_{m}(x,u)\tilde N_{m}(y,v)\nn\\
&\quad\quad\times\left[\frac{\partial^2\Gamma_k}{\partial\vp(u)\partial\vp(v)}+\int_zC(u,v,z)\frac{\partial\Gamma_k}{\partial\vp(z)}\right]\
,
\end{align}
with \qq{connection}
\bel{F26}
C(u,v,z)=\int_{s,w}\frac{\partial\chi(s)}{\partial\vp(u)}
\frac{\partial\chi(w)}{\partial\vp(v)}
\frac{\partial^2\vp(z)}{\partial\chi(s)\partial\chi(w)}\ .
\ee
This connection results in a correction term,
\bel{F27}
\Delta(u,v)=\int_zC(u,v,z)\frac{\partial\Gamma_k}{\partial\vp(z)}\ ,
\ee
which appears in the flow equation~\eqref{F16} since $\partial_t\Gamma|_\chi$
involves $\partial^2\Gamma/\gl\partial\chi(x)\partial\chi(y)\gr$.

For a local field transformation, $\partial\vp(x)/\partial\chi(y)=\tilde
N_{m}(y)\delta(x-y)$, the connection reads
\begin{align}
\label{F28}
C(u,v,z)=&\,-\frac{\partial\tilde
N_{m}^{-1}}{\partial\chi}(u)\delta(u-v)\delta(u-z)\ ,\nn\\
\Delta(u,v)=&\,-\frac{\partial\tilde
N_{m}^{-1}}{\partial\chi}(u)\frac{\partial\Gamma_k}{\partial\vp(u)}\delta(u-v)\
.
\end{align}
With
\bel{F29}
\tilde\cR_k(u,v)=\int_{s,w}\tilde N_{m}^{-1}(u,s)\tilde
N_{m}^{-1}(v,w)\cR_k(s,w)
\ee
one obtains the flow equation
\begin{align}
\label{F30}
\partial_t\Gamma_{\chi,k}[\vp]=\frac12\tr&\left\{\tilde
N_{m}^{-1}\partial_t\cR_k\gl\tilde
N_{m}^{-1}\gr^T\gl\Gamma_{(\vp)}^{(2)}+\Delta+\tilde\cR_k\gr^{-1}\right\}\nn\\
&-\int_x\frac{\partial\Gamma_k}{\partial\vp(x)}\partial_t\vp(x)|_\chi\ .
\end{align}
The second functional derivative with respect to $\vp$,
$\Gamma_{(\vp)}^{(2)}(x,y)=\partial^2\Gamma_k/\partial\vp(x)\partial\vp(y)$, is
effectively regulated by $\tilde\cR_k$ which becomes field-dependent. Even if
one finds a field transformation which brings $\Gamma_{(\vp)}^{(2)}+\Delta$ to a
simple form, there remains an additional complication from the field dependence
of $\tilde\cR_k$. The cutoff acts no longer homogeneously for all field values.

The field transformation discussed in this section only involves the macroscopic
fields. The effective average action and the flow equation is not changed, they
are only expressed in terms of different variables. The same holds for the field
equations that follow in the absence of sources,
$\partial\Gamma_k/\partial\chi(x)=0$ or $\partial\Gamma_k/\partial\vp(x)=0$.
These differential equations are simply expressed in different variables. For a
non-linear field transformation there is no direct relation between the
\qq{composite macroscopic field} $\vp$ and the expectation value of some
microscopic composite field. For $\vp(x)=f\gl\chi(x)\gr$ the relation
$\vp(x)=\big\langle f\gl\chi'(x)\gr\big\rangle$ does not hold, except for linear
functions $f$. In this sense the canonical field $\chi(x)$ is singled out by the
relation $\chi(x)=\langle\chi'(x)\rangle$.

\subsection*{Elimination of couplings}

Field transformations of the macroscopic fields can be used to bring either the
flow equation or the field equation into a form easier to solve approximately or
simpler to understand. An example for the transformation of the flow equation is
the elimination of certain couplings~\cite{GW1, GW2, BRA, BBW, SAL} or
simplification of coupling functions~\cite{FKW, DMP}. A given flowing coupling
is typically defined by some functional derivative of $\Gamma_k$. Assume a
setting where this coupling vanishes for the \qq{microscopic effective action}
$\Gamma_{k=\Lambda}$ with very large (or infinite) $\Lambda$. The flow of this
coupling obtains by a suitable functional derivative of the flow
equation~\eqref{F30}. In general, the coupling will be generated by the flow.
One can now choose a field transformation defined by the property that the flow
of this particular coupling vanishes. This poses a constraint on the choice of
$\vp_k[\chi]$. This procedure is useful provided that for this choice of
$\vp_k[\chi]$ no other important unwanted couplings are generated.

As an example we consider a truncation of the effective average action similar
to eq.~\eqref{M1}
\bel{F31}
\Gamma_{\chi,k}[\chi]=\frac12\int_xK(\chi)\partial^\mu\chi\partial_\mu\chi\ ,
\ee
where $K(\chi)$ is a $k$-dependent function of $\chi$. With the non-linear field
transformation $\vp[\chi]$ defined by
\bel{F32}
\frac{\partial\vp}{\partial\chi}=\tilde N_{m}=K^{1/2}\ ,\quad
\Gamma_{\chi,k}[\vp]=\frac12\int_x\partial^\mu\vp\partial_\mu\vp\ ,
\ee
the effective average action is the one for a free massless scalar field.
As long as $K_k(\chi)$ is positive for all $\chi$ this field transformation
exists for every $k$. Naively, one could conclude that fluctuations of a free
field do not generate interactions or a scalar potential and that for every $k$
we deal with a free theory for composite fields.

In order to see that this is not correct in general, we investigate the flow
equation with a suitable macroscopic variable transformation which keeps the
kinetic term~\eqref{F32} fixed for all $k$. This will also fix the
$k$-dependence of $K_k(\chi)$. For the modified flow equation~\eqref{F30} one
employs
\begin{align}
\label{F33}
\Gamma_{(\vp)}^{(2)}=&\,-\partial^\mu\partial_\mu=-\partial^2\ ,\quad
\frac{\partial\Gamma_k}{\partial\vp(x)}=-\partial^\mu\partial_\mu\vp(x)\ ,\nn\\
\Delta=&\,-\frac{\partial K^{-1/2}}{\partial\chi}\partial^\mu\partial_\mu\vp\ .
\end{align}
With
\bel{106A}
\vp=N_{m}\chi\ ,\quad \partial_t\vp=\partial_tN_{m}\chi=\partial_t\ln N_{m}\vp\
,
\ee
the last term in eq.~\eqref{F30} becomes
\begin{align}
\label{106B}
-\int_x\frac{\partial\Gamma}{\partial\vp(x)}\partial_t\vp(x)=&\,
-\int_xA(\vp)\partial^\mu\vp\partial_\mu\vp\ ,\nn\\
A(\vp)=&\,\frac{\partial}{\partial\vp}\gl\partial_tN_{m}\chi\gr\ ,
\end{align}
where $\partial_tN_{m}$ is evaluated for fixed $\chi$.

Extracting from a derivative expansion of the first term on the r.h.s. of
eq.~\eqref{F30} the coefficient $B(\vp)$ of the term with two derivatives yields
\bel{106C}
\partial_t\Gamma[\vp]=\int_x\big[B(\vp)-
A(\vp)\big]\partial^\mu\vp\partial_\mu\vp+\dots
\ee
One can now determine $\partial_t(N_{m}\chi)$ such that $A(\vp)=B(\vp)$,
\begin{align}
\label{106D}
\frac{\partial}{\partial\vp}\partial_t(N_{m}\chi)=&\,B(\vp)\ ,\nn\\
\partial_tN_{m}(k,\chi)=&\,
\frac1\chi\int_{\chi_0}^\chi\text{d}\chi'\left(N_{m}(\chi')+\frac{\partial
N_{m}(\chi')}{\partial\ln\chi'}\right)B(\chi')\ .
\end{align}
In this case the canonical kinetic term~\eqref{F32} for $\vp$ remains preserved.
In other words, possible couplings $\sim\vp^2\partial^\mu\vp\partial_\mu\vp$
etc. are eliminated by this $k$-dependent macroscopic field transformation.

Nevertheless, the flow equation with initial condition~\eqref{F31} for
$k=\Lambda$ has not a free massless scalar field as a solution. The dots in
eq.~\eqref{106C} indicate terms in the effective average action with zero or
more than two derivatives. These terms are generated by the flow even if not
present in the microscopic effective action. For example, the flow equation for
a non-trivial potential obtains by evaluating eq.~\eqref{F30} for homogeneous
$\vp$. If not present at $\Lambda$, this flow generates a potential for
$k<\Lambda$. For a non-linear field transformation the r.h.s. of eq.~\eqref{F30}
does not vanish due to the $\vp$-dependence of $\tilde N_{m}$.

If one has found $\partial_tN_{m}(k,\chi)$ as a solution of eq.~\eqref{106D} one
can
infer for the flow of the kinetial $K(\chi)$ the relation
\bel{106E}
\partial_tK^{1/2}(k,\chi)=
\left(1+\frac{\partial}{\partial\ln\chi}\right)\partial_tN_{m}(k,\chi)\ .
\ee
This can be compared with the flow of the kinetial computed without a field
transformation in appendix~\ref{app:FEK}. Without a truncation the two
computations should yield identical results for $K(k,\chi)$. The effective
actions $\Gamma_{\chi,k}[\chi]$ and $\Gamma_{\chi,k}[\vp]$ are related by a
simple variable transformation. This $k$-dependent macroscopic field
transformation also maps the flow equation for $\partial_t\Gamma|_{\chi}$ and
$\partial_t\Gamma|_\vp$ into each other. If a kinetial $K(k,\chi)$ obeys the
flow equation at fixed $\chi$, the field transformation is constructed such that
the kinetic term for $\vp$ takes for all $k$ the simple form~\eqref{F32}. The
kinetial in the $\vp$-scheme does not change. For the truncation~\eqref{F31}
this is precisely the condition~\eqref{106D}. For the present example the direct
method of appendix~\ref{app:FEK} seems actually to be much simpler than the use
of a field transformation.

It is sometimes argued that couplings contained in the kinetial $K(\chi)$ are
\qq{inessential} since they can be removed by a field transformation which leads
to a canonical kinetic term. Field transformations only move, however, the
information contained in the couplings from one place to another. Only in the
truncation~\eqref{F31} the precise form of $K(\chi)$ does not matter since this
action describes a free massless composite scalar field $\vp$ for arbitrary
monotonic $K(\chi)$. As soon as one computes the flow of other parts in the
effective action, in particular the flow of the potential, the information
contained in $K(\chi)$ becomes important. In appendix~\ref{app:FKP} we
demonstrate explicitly the importance of the potential for the flow of the
kinetial $K$. One concludes that the omission of the dots in eq.~\eqref{106C}
can be strongly misleading. We are aware that omitting the potential is a rather
drastic truncation. Our example is thought to underline, however, a general
issue for field transformations in truncated flow equations. One has to ensure
that the field transformation does not move important information into parts
that are not covered by the truncation.

\section{Flow equations with cutoff and sources for composite
fields}\label{sec:FECS}

In this section we investigate a different form of field-transformations in flow
equations. In this version the variable transformation is performed on the level
of the precise definition of the effective average action by choosing a
different source term and a different form of the infrared cutoff, now adapted
to composite fields. Since the field transformation leads to a different
functional integral we call it a microscopic field transformation, in
distinction to the macroscopic field transformation which only changes the
arguments of the effective average action, as discussed in the preceding
section. Microscopic field transformations are closely related to the
possibility of defining different effective actions, as $\Gamma_\chi$ or
$\Gamma_\vp$ in sect.~\ref{sec:FTEA}. They extend this concept to non-zero $k$.
We will see in sect.~\ref{sec:TFF} that for the microscopic field transformation
the Jacobian reappears in the initial value for the flow equations at very high
$k$. This happens independently of a change of variables for the functional
measure or not. The appearance of the Jacobian underlines the close connection
of the microscopic field transformations to the microscopic field
transformations in the functional integral in sect.~\ref{sec:FTFI}. The
microscopic field transformations discussed in this section have been derived in
ref.~\cite{PAW}. They correspond to the ones employed for the \qq{essential
renormalization group}~\cite{BZF, BAFA}

\subsection*{Effective average action for composite fields}

As an alternative to the effective average action $\Gamma_{\chi,k}$ defined in
the preceding section, we define a different version of the effective average
action $\Gamma_{\vp,k}$ by taking the cutoff piece quadratic in the fields
$\vp'_k[\chi']$ and using a source given by
$\partial\Gamma'_{\vp,k}/\partial\vp(x)$,
\begin{align}
\label{C1}
&\Gamma'_{\vp,k}[\vp]=-\ln\int\cD\chi'\exp\bigg\{-S[\chi']\\
&-\frac12\int_{x,y}\vp'_k(x)\cR_k(x,y)\vp'_k(y)+\int_x\frac{\partial\Gamma_{\vp,k}'}{\partial\vp(x)}\gl\vp'_k(x)-\vp(x)\gr\bigg\}\
.\nn
\end{align}
Here $\vp'_k(x)$ is a functional of $\chi'$, $\vp'_k[\chi'](x)$ which can depend
on $k$. For the effective action we subtract again the cutoff piece
\bel{C2}
\Gamma_{\vp,k}[\vp]=\Gamma'_{\vp,k}[\vp]-\frac12\int_{x,y}\vp(x)\cR_k(x,y)\vp(y)\
.
\ee
For $k=0$ and $\cR_{k=0}=0$ this effective average action becomes the quantum
effective action $\Gamma_\vp[\vp]$, as defined by eq.~\eqref{E27A}. As discussed
in section~\ref{sec:FTEA}, the effective actions $\Gamma_\vp[\vp]$ and
$\Gamma_\chi[\vp]$ are not identical, due to the different form of the source
term. Thus $\Gamma_{\chi,k}$ and $\Gamma_{\vp,k}$ differ for all $k$, including
$k=0$. An exception are the values for solutions of the field equations which
coincide for $k=0$.

We emphasize that the definition~\eqref{C1},~\eqref{C2} keeps the original
measure $\cD\chi'$ and the classical action $S[\chi']$ unchanged. There is no
need to switch to $k$-dependent integration variables $\vp'_k(x)$, which would
introduce a $k$-dependent Jacobian and make the classical action $k$-dependent
once expressed as a functional of $\vp'_k$. We could, however, use different
integration variables $\vp'$. In this form $\Gamma_{\vp,k}$ takes a form
analogous to eq.~\eqref{F2} provided we include the Jacobian by replacing the
microscopic action $S[\chi']$ by $\Sbar[\vp']$. This gives a hint why the
Jacobian will be back for the initial conditions of $\Gamma_{\vp,k}$ even if we
do not change the integration variable. With the integration variable $\vp'$ the
flow equation evaluated at fixed $\vp$ will receive an additional contribution
from the $k$-dependence of the Jacobian or $\Sbar[\vp']$. On the other hand,
keeping the original integration variables $\chi'$ the flow equation will
receive additional contributions from the $k$-dependence of the variable
transformation. The two views are equivalent.

Taking the functional derivative of $\Gamma_{\vp,k}$ with respect to $\vp(x)$
establishes similar to eqs.~\eqref{E4},~\eqref{E7} the identity
\bel{C3}
\langle\vp'(x)\rangle=\vp(x)\ ,
\ee
where the expectation value is taken in the presence of the source
$\partial\Gamma'_{\vp,k}/\partial\vp(x)$ and the cutoff term $\sim\cR_k$.
Similarly, the derivative of eq.~\eqref{C3} with respect to $\vp(y)$ yields an
identity corresponding to eq.~\eqref{E8}
\bel{C4}
\int_y\frac{\partial^2\Gamma_{\vp,k}'}{\partial\vp(x)\partial\vp(y)}\gl\langle\vp'_k(y)\vp'_k(z)\rangle-\vp(y)\vp(z)\gr=\delta(x-z)\
.
\ee

\subsection*{Flow equation for \qq{composite} cutoff and sources}

The flow equation receives an additional contribution from the $k$-dependence of
$\vp'_k[\chi]$,
\begin{align}
\label{C5}
\partial_k\Gamma_{\vp,k}[\vp]=&\,\frac12\tr\Big\{\partial_k\cR_k\gl\Gamma_{\vp,k}^{(2)}+\cR_k\gr^{-1}\Big\}\nn\\
&-\int_x\frac{\partial\Gamma_{\vp,k}}{\partial\vp(x)}\langle\partial_k\vp'_k(x)\rangle\\
&+\int_{x,y}\cR_k(x,y)\big\langle\gl\vp'_k(y)-\vp(y)\gr\partial_k\vp'_k(x)\big\rangle\
.\nn
\end{align}
The $k$-derivative of $\vp'_k$ is taken at fixed $\chi'$. The additional terms
in the flow equation involve $\partial_k\vp'_k$ and therefore vanish for a
$k$-independent field transformation. Expressed in terms of $\chi'$ the
additional terms involve higher correlation functions.

For the last term in eq.~\eqref{C5} we can employ a general identity for
operators $\cA[\vp']$,
\bel{C15}
\big\langle\gl\vp'(x)-\vp(x)\gr\cA[\vp']\big\rangle=\int_yG(x,y)\frac{\partial}{\partial\vp(y)}\langle\cA[\vp']\rangle\
,
\ee
with $G(x,y)=\gl\Gamma_{\vp,k}^{(2)}+\cR_k\gr^{-1}(x,y)$ given by
eq.~\eqref{F6}. This is found directly by using the definition of the
expectation value in the presence of sources or the associated value of $\vp$,
and the infrared cutoff term
\bel{C16}
\cA[\vp]=\langle\cA[\vp']\rangle=\frac{\int\cD\chi'\cA[\vp']\exp\big\{-F[\chi']\big\}}{\int\cD\chi'\exp\big\{-F[\chi']\big\}}\
,
\ee
with
\bel{C16A}
F[\chi']=S[\chi']+\Delta_kS[\chi']-\int_y\frac{\partial\Gamma'_{\vp,k}}{\partial\vp(y)}\gl\vp'(y)-\vp(y)\gr\
.
\ee
Here $\Delta_kS$ involves the cutoff term (second term in the curled bracket in
eq.~\eqref{C1}), which depends on $\chi'$ through $\vp'(\chi')$. Transporting
the $\vp$-derivative inside the integral,
\bel{C17}
\frac{\partial}{\partial\vp(x)}\cA[\vp]=\int_y
\frac{\partial^2\tilde\Gamma}{\partial\vp(x)\partial\vp(y)}
\big\langle\gl\vp'(y)-\vp(y)\gr\cA[\vp']\rangle\ ,
\ee
and multiplication with $G$ yields eq.~\eqref{C15}.

Inserting for $\cA[\vp']$ the logarithmic $k$-derivative of $\vp'_k(x)$, with
expectation value
\bel{C18}
\gamma(x)=\langle\partial_t\vp'_k(x)\rangle\ ,
\ee
one arrives at the flow equation
\begin{align}
\label{C19}
\partial_t\Gamma_{\vp,k}=&\,\frac12\int_{x,y}\bigg[\partial_t\cR_k(x,y)+2\int_z\frac{\partial\gamma(z)}{\partial\vp(x)}\cR_k(z,y)\bigg]\nn\\
&\times\gl\Gamma_{\vp,k}^{(2)}+\cR_k\gr^{-1}(y,x)-\int_x\frac{\partial\Gamma_{\vp,k}}{\partial\vp(x)}\gamma(x)\
.
\end{align}
This is the form of the flow equation discussed in refs.~\cite{PAW, BZF, BAFA,
IPA}, in analogy to the discussion of frame independence by Wegner~\cite{WEG},
see also~\cite{SAL}. The quantity $\gamma(x)$ is often called the
\qq{RG-kernel}.

This flow equation shows similarities with the flow equation~\eqref{F30} for
$\Gamma_{\chi,k}[\vp]$. For linear field transformations the macroscopic and
microscopic field transformations lead to identical results. For the macroscopic
field transformations the connection term vanishes. For the microscopic field
transformations the term involving $\partial\gamma/\partial\vp$ does not depend
on $\vp$. The flow equation~\eqref{C19} coincides with the flow
equation~\eqref{F16} provided that one chooses a suitable adapted cutoff
$\cR_k$. This may be verified by reproducing eq.~\eqref{F23} for the
transformation~\eqref{F21}. For eq.~\eqref{F17} one has
$\tilde\gamma_k[\vp]=\gamma_k[\vp]$.

For non-linear field transformations one finds some important differences. As
may be expected, the main difference concerns the infrared cutoff, which is now
a bilinear form in the field $\vp'$ instead of $\chi'$. As a consequence, there
is no \qq{connection term} $\Delta$ as in eq.~\eqref{F30}. Also the factors
$\tilde N_{m}^{-1}$ reflecting the \qq{inhomogeneity} of the cutoff translated
to
the composite field are absent. The \qq{correction terms} involving $\gamma$ in
eq.~\eqref{C19} have a somewhat different form as compared to eq.~\eqref{F30}.

We observe that the relation between the source and the macroscopic composite
field $\vp$ depends on $k$. For a fixed source the macroscopic field depends on
$k$. Different prescriptions on what is kept fixed during the flow lead to
different versions of flow equations even if the same cutoff function for the
composite field $\vp'$ is used. An early version of a flow equation with cutoff
for \qq{composite} fields $\vp'$ can be found in ref.~\cite{GW1}. All the
different versions of field transformations and corresponding definitions of
$\Gamma_k[\vp]$ have been used in practice for a simplification of flow
equations, for example by absorbing unwanted terms in the effective action by a
suitable choice of $k$-dependent \qq{composite fields}. This can be achieved by
the \qq{correction terms} induced by $k$-dependent fields canceling terms
induced
by the trace term in the flow equation. Flowing bosonization is a good example
of this concept~\cite{GW2, BRA}. Also the ideas of \qq{essential renormalization
flow}~\cite{BZF, BAFA} or \qq{optimized flow}~\cite{IPA} rely on the elimination
of couplings in the effective average action by a suitable choice of field
variables.

\subsection*{RG-kernel}

The version with infrared cutoff and source for composite fields offers for
non-linear transformations an important advantage as compared to
eq.~\eqref{F30}: The cutoff can be adapted to the propagator for composite
fields. The inhomogeneity of the cutoff due to the field-dependent factors
$\tilde N_{m}^{-1}$ can be avoided. There is, however, a price to be paid for
this setting. First, the RG-kernel is, in general, not easily computable for a
given field transformation. Inversely, for a given RG-kernel the underlying
microscopic field transformation is often not known. Second, for non-linear
microscopic field transformations the direct contact of the effective action for
large $k$ to the classical action is lost. We address the second issue in the
next section and discuss here some consequences of the non-linearity of the
field transformation.

For $\vp'_k$ depending non-linearly on $\chi'$ the $k$-derivative typically
involves higher correlation functions of $\vp'_k$. As an example, we take
\bel{C6}
\vp'_k(x)=\chi'(x)+g(k)\chi'(x)^2\ ,
\ee
with
\bel{C7}
\partial_k\vp'_k(x)=\partial_kg(k)\chi'(x)^2=\partial_k\ln
g(k)\gl\vp'_k-\chi'(\vp'_k)\gr\ ,
\ee
where
\bel{C8}
\chi'(\vp'_k)=\frac{1}{2g}\gl\sqrt{1+4g\vp'_k}-1\gr\ .
\ee
Expanding around some arbitrary $\vp'_{k0}$, arbitrarily high powers of $\vp'_k$
appear in $\partial_k\vp'_k(x)$. This results for the flow
equation~\eqref{C5},~\eqref{C19} in the presence of correlation functions for
$\vp_k'$ of arbitrary high order. Those could be computed, in principle, by
expressions involving high functional derivatives. This destroys, however, the
simple closed form of the flow equation. At this point the $k$-dependence of
$\vp'_k$ reflects only the choice of composite field $\vp'_k[\chi']$. In
general, this choice needs not to be related to the flow of the effective
action.

One may go the opposite way and specify the RG-kernel $\gamma$ such that the
flow equations simplify. In this case the microscopic transformation
$\vp_k'(\chi')$ is often no longer known and the existence of an invertible
transformation of the microscopic variables is not guaranteed. One may try to
construct the microscopic field transformation $\vp'_k[\chi']$ from a given
$\gamma_k[\vp]$. For this purpose we integrate the equation,
\bel{126A}
k\partial_k\vp(k;\chi)=\gamma_k\big[\vp(k;\chi)\big]\ ,
\ee
with boundary condition
\bel{126B}
\vp(\Lambda;\chi)=\chi\ .
\ee
One may take large $\Lambda$ associated to a scale in microphysics or
$\Lambda\to\infty$. The solution of this differential equation $\vp(k;\chi)$
describes how the expectation value $\langle\vp'_k(\chi')\rangle=\vp(k;\chi)$
evolves with $k$ if for $k=\Lambda$ one has
$\langle\vp'_\Lambda(\chi')\rangle=\chi$.

Eq.~\eqref{126A} and its solution concerns the expectation value and needs
translation to $\vp'_k(\chi')$. A microscopic field transformation
$\vp'_k(\chi')$ with $\vp'_\Lambda(\chi')=\chi'$ obeys
$\langle\vp'_\Lambda(\chi')\rangle=\langle\chi'\rangle$ and we can choose
$\chi=\langle\chi'\rangle$. One may try to use for all $k$ the ansatz
\bel{128A}
\vp'_k(\chi')=\vp(k;\chi')\ ,
\ee
with $\vp(k;\chi)$ the solution of eqs.~\eqref{126A},~\eqref{126B}. Evaluating
the $k$-derivative of the expectation value of eq.~\eqref{128A} yields
\bel{128B}
\langle\partial_t\vp'_k(\chi')\rangle=\big\langle\gamma_k\big[\vp'_k(\chi')\big]\big\rangle\
.
\ee
On the other hand, the definition of the RG-kernel reads
\bel{128C}
\langle\partial_t\vp'_k(\chi')\rangle=\gamma_k\big[\langle\vp'_k(\chi')\rangle\big]\
.
\ee
Eqs.~\eqref{128B} and~\eqref{128C} are compatible only for linear RG-kernels. If
$\gamma_k$ depends non-linearly on $\vp$ the ansatz~\eqref{128A} leads to a
contradiction since
$\langle(\vp'_k(\chi'))^n\rangle\neq\langle\vp'_k(\chi')\rangle^n$. For
non-linear RG-kernels this type of reconstruction does not work. We see no
simple way to find the microscopic field transformation $\vp'_k(\chi')$ for a
given non-linear $\gamma_k[\vp]$.

If one is able to find the solution $\vp(k;\chi)$ for a linear RG-kernel, one
can then judge if $\vp'_k(\chi')$ remains invertible for all $k$. In contrast,
for a non-linear RG-kernel it seems to be very hard in practice to verify a
given $\gamma_k[\vp]$ corresponds to a regular microscopic field transformation
$\vp'_k(\chi')$.

\subsection*{Linear RG-kernel}

For linear RG-kernels $\partial_t\vp'_k$ is linear in $\vp'_k$. This can be
realized for a particular family of non-linear field transformations
$\vp'_k(\chi')$.

Let us consider local variable transformations of the form
\bel{C20}
\vp'=N(\chi')\chi'\ ,
\ee
with RG-kernel
\bel{C21}
\gamma(x)=\big\langle\gl\partial_tN(\chi')\chi'\gr(x)\big\rangle\ .
\ee
For general functions $N(\chi')$ there is no simple expression for $\gamma(x)$.
For the special case
\bel{C22}
\partial_tN(\chi')=a(k)N(\chi')+\frac{b(k)}{\chi'}\ ,\quad
\partial_t\vp'=a\vp'+b\ ,
\ee
the expression $\partial_t\vp'=\partial_tN(\chi')\chi'$ is linear in $\vp'$ and
therefore implies a linear RG-kernel
\bel{C23}
\gamma(x)=a(k)\vp(x)+b(k)\ .
\ee
This particular choice of the field $\vp'(\chi')$ yields the simple flow
equation
\begin{align}
\label{C24}
\partial_t\Gamma_{\vp,k}=&\,\frac12\tr\Big\{\gl\partial_t+2a(k)\gr\cR_k\gl\Gamma_{\vp,k}^{(2)}+\cR_k\gr^{-1}\Big\}\nn\\
&-\int_x\frac{\partial\Gamma_{\vp,k}}{\partial\vp(x)}\gl a(k)\vp(x)+b(k)\gr\ .
\end{align}

In the appendix~\ref{app:FCSF} we employ microscopic field transformations with
a linear RG-kernel in order to discuss under which circumstances the flowing
effective action~\eqref{F31} yields a model of free composite scalar fields. We
conclude that using linear RG-kernels~\eqref{C23} the correction terms for
non-zero $a$ or $b$ in the flow equation~\eqref{C24} can be employed to achieve
certain simplifications. For this restricted set of field transformations one
can keep control of the microscopic transformations.

This is no longer the case for more general non-linear RG-kernels $\gamma(\vp)$
to which we turn next. The possibility to assume a suitable non-linear RG-kernel
without explicit knowledge of the microscopic field transformation is further
discussed in sect.~\ref{sec:SS}.

\subsection*{Power and dangers of non-linear microscopic field transformations}

Microscopic field transformations with a non-linear RG-kernel offer very rich
possibilities to change the functional form of the effective average action and
associated flow equations. This may be regarded as a powerful tool for a
simplification of the flow equations. Terms in the flow equations which are
responsible for technical difficulties can be removed by a suitable choice of
the RG-kernel. Some proponents of the essential renormalization group suggest
that all couplings that can be removed by a suitable choice of a non-linear
RG-kernel can be regarded as redundant and play no role for observations. For
example, it has been argued that in quantum gravity the cosmological constant is
redundant in this sense.

This view is dangerous in two aspects. First, it is often not guaranteed that a
given choice of a non-linear RG-kernel $\gamma[\vp]$ can indeed be realized by a
regular microscopic field transformation. We have seen that in general the
relation between the RG-kernel and the field transformation is rather complex.
In many cases it is not accessible in practice. The requirement of regularity of
the microscopic field transformation imposes constraints on the allowed form of
$\gamma[\vp]$. Omitting these constraints which are unknown in practice presents
then the danger that the chosen form of $\gamma[\vp]$ is actually not consistent
with a microscopic definition of the effective average action.

The second danger is related to the necessity of truncations. In practice, this
is often the most important shortcoming. Assuming that a regular field
transformation producing a given RG-kernel exists, the flow equation~\eqref{C19}
is an exact identity for a well defined effective action which generates
correlation functions for composite fields. A solution of this flow equation
requires in most cases a truncation of the general form of the effective average
action. The danger of non-linear microscopic field transformations consists in
the problem that important physical information can be moved to sectors which
are no longer resolved for a given truncation.

As an example we may again consider the momentum dependence of the propagator
for electrons in a solid. It contains central information on the Fermi surface
and many central quantities for observations. The field dependence of this
propagator encodes key aspects as the possibility of spontaneous symmetry
breaking. By a suitable choice of the RG-kernel one may transform this momentum-
and field-dependence to a simple standard form which no longer reflects the
details of the Fermi surface. As discussed in the introduction, this does not
mean that the Fermi surface or the effects of spontaneous symmetry breaking on
the fermion propagator are not observable. The field transformation only shifts
the information about the physical Fermi surface to higher order correlation
functions. If a truncation omits the detailed momentum- and field-dependence of
vertices, the use of the field transformation can remove key physical
information by the truncation.

In the context of the simple scalar field theory we will demonstrate this issue
by the role of the effective scalar potential. The use of suitable RG-kernels
$\gamma[\vp]$ can both produce and remove the effective potential in the
effective average action. Removing the potential by a microscopic field
transformation shifts the relevant physical information to sectors that are more
difficult to access. A too simple truncation of these sectors will discard
important physical information.

\subsection*{Effective scalar potential}

The linear RG-kernel~\eqref{C23} is special in the sense that no scalar
potential for $\vp$ is generated. For non-linear kernels the derivative
$\partial\gamma/\partial\vp$ will depend on $\vp$. Then the flow generates a
potential even if one starts at high $k$ without a potential. We demonstrate
this for a setting with constant kinetial $K=Z$.

The flow equation for the effective potential is found by evaluating the flow of
$\Gamma_k$ for a homogeneous scalar field $\vp(x)=\vp$,
$\Gamma_k=\int_xU_k(\vp)$. We assume that in this case also $\gamma$ is
homogeneous, $\gamma=\gamma(\vp)$. The flow equation for $U_k$ obtains from
eq.~\eqref{C19} as
\begin{align}
\label{C49}
\partial_tU_k=&\,\frac12\int_q\left[\partial_t\cR_k(q^2)+2\frac{\partial\gamma}{\partial\vp}\cR_k(q^2)\right]\big[Zq^2+\cR_k(q^2)\big]^{-1}\nn\\
&+\Delta\partial_tU\ .
\end{align}
Here $\Delta\partial_tU$ indicates terms proportional to derivatives of $U$ with
respect to $\vp$ or contributions from fluctuations of other fields. As long as
these derivatives are small we can employ on the r.h.s. of the flow equation
$\Gamma_k=(Z/2)\int_x\partial^\mu\vp\partial_\mu\vp$. The $\vp$-dependent part
of the potential therefore flows as
\bel{C50}
\partial_t\left(\frac{\partial
U}{\partial\vp}\right)=\frac{\partial^2\gamma}{\partial\vp^2}\int_q\frac{\cR_k(q^2)}{Zq^2+\cR_k(q^2)}=C
k^4\frac{\partial^2\gamma}{\partial\vp^2}\ .
\ee

If we start at very high $k=\Lambda$ with a flat potential, $\partial
U/\partial\vp=0$, a non-zero slope will develop for $k<\Lambda$. For positive
$\partial^2\gamma/\partial\vp^2>0$ this slope is negative and leads to a
tendency towards spontaneous symmetry breaking. For negative
$\partial^2\gamma/\partial\vp^2<0$ the potential develops typically a minimum at
$\vp=0$ with a positive mass term $m_\vp^2\sim
k^4\partial^2\gamma/\partial\vp^2$. (We assume here a discrete symmetry
$\vp\to-\vp$ for which $\gamma(-\vp)=-\gamma(\vp)$,
$\gamma(\vp)=a\vp+d\vp^3+\hdots$.)

One could also be tempted to use the non-linear field transformations in order
to remove the effective scalar potential. Imagine that the fluctuations of
additional fields, say fermions with a Yukawa coupling to the scalar, induce on
the r.h.s. of eq.~\eqref{C49} terms of the type
\bel{161A}
\Delta\partial_tU=A_1k^2\vp^2+A_2\vp^4\ .
\ee
One could then take an RG-kernel of the form
\bel{161B}
\gamma=-\frac{A_1\vp^3}{3Ck^2}-\frac{A_2\vp^5}{5Ck^4}\ ,
\ee
such that the combined flow of $\partial_t(\partial U/\partial\vp)$ vanishes. If
a microscopic field transformation leading to the RG-kernel~\eqref{161B} exists,
the potential for $\vp$ remains flat for all $k$. In the language of the
\qq{essential renormalization group} it would seem that one can remove the
potential as \qq{irrelevant couplings}.

\subsection*{Conservation of physical information}

The possibility to remove the effective potential by a non-linear field
transformation may seem surprising at first sight. Fermion fluctuations
typically induce terms of the form
\bel{161C}
A_1=\frac{a_1k^2}{k^2+m^2}\ ,\quad A_2=-\frac{a_2k^4}{(k^2+m^2)^2}\ ,
\ee
with $m$ the mass of the fermions depending on $\vp$ via a Yukawa coupling.
Without removing the potential by the field transformation~\eqref{161B} the
fermion fluctuations would induce for $k\to0$ a potential indicating spontaneous
symmetry breaking,
\bel{161D}
U_{k\to0}=-\frac{\mu^2}{2}\vp^2+\frac\lambda8\vp^4+\dots\ .
\ee
(We assume a vanishing potential at some microscopic scale $\Lambda$.) The
question arises where to see this spontaneous symmetry breaking after the
potential has been eliminated by a field transformation.

Field transformations can only move physical features to other sectors. On the
one side the last term in the flow equation~\eqref{C19} generates a more complex
scalar kinetic term according to
\bel{161E}
\int_x\partial^\mu\partial_\mu\vp\gamma(\vp)=-\int_x\frac{\partial\gamma}{\partial\vp}\partial^\mu\vp\partial_\mu\vp\
.
\ee
On the other side, this term also induces more complex interactions between the
scalar and fermions $\psi$ (with $h$ a Yukawa coupling) according to
\bel{161F}
\int_x-\gamma(x)\frac{\partial}{\partial\vp(x)}\int_yh\vp(y)\psibar(y)\psi(y)=-\int_xh\gamma(\vp)\psibar\psi\
.
\ee
The physical information about spontaneous symmetry breaking induced by fermion
fluctuation is only lost if a truncation discards these modifications of the
kinetial and Yukawa coupling. Only if a truncation keeps the field-dependence of
the kinetial and scalar-fermion coupling one can recover the spontaneous
symmetry breaking even for a field choice that eliminates the effective
potential.

It may be rather difficult to extract the features of spontaneous symmetry
breaking from the flow of these higher correlation functions. Given the
simplicity of the picture with a scalar potential it seems preferable to keep as
much of the information as possible in the form of the scalar potential. Similar
considerations apply to the physical information about field-dependent
structures in the propagator or two-point function, as a gap induced by
spontaneous symmetry breaking. While non-linear field transformations may allow
one to move the information contained in the two-point function to other
sectors, it will be, in general, more cumbersome to extract the relevant
information from those higher correlation functions.

At this point it may be useful to highlight an important difference between
macroscopic and microscopic field transformations. The macroscopic field
transformations do not change the effective average action. It only expresses a
given functional in terms of different field-variables. In this way one can
explicitly follow the action of the transformation once a given approximation to
$\Gamma_k$ is given. For example, a transformation
$\vp(x)=f^{-1}_k\gl\chi(x)\gr$ transforms $U(\chi)\to U(\vp)=U\gl f_k(\vp)\gr$.
A regular transformation $f_k$ can transform an arbitrary convex potential
$U(\chi)$ to $U(\vp)=m^2\vp^2+U_0$ for a suitable range of fields. This holds
for arbitrarily small $m^2$. A potential with spontaneous symmetry breaking of
the discrete symmetry $\chi\to-\chi$ cannot be transformed to this form,
however. Monotonically decreasing potentials can be brought to the form
$U=M^4\exp(-\vp/M)$ for a suitable range of fields. Transforming the potential
to a standard form shifts information to the kinetial. Standard forms of the
potential have proven to be useful for inflationary cosmology~\cite{CWCI, KLR,
Wetterich_2015, GKL} or dynamical dark energy~\cite{HEWE, CWCQ}.

Microscopic field transformations change the flow equation and the effective
average action beyond the simple use of different variables. Their action is
much less under control and it is often not obvious to which sector the relevant
information is shifted. For example, one may doubt if the non-linear RG-kernel
which removes the scalar potential completely can be realized by a suitable
regular field transformation.

\section{Flow equations for fundamental and composite fields}\label{sec:TFF}

Two issues remain to be settled if one uses the IR-cutoff for composite fields.
The first concerns the relation between the microscopic action $S[\chi']$ and
the microscopic effective action $\Gamma_{\vp,k}$ for very large $k=\Lambda$, or
for $k\to\infty$. The second involves the relation between $\Gamma_{\vp,k}[\vp]$
and expectation values and correlations for the \qq{fundamental field} $\chi'$.
In particular one would like to know the relation between $\langle\chi'\rangle$
and $\vp$. Knowing $\langle\chi'\rangle=F[\vp]$ one could employ a macroscopic
variable transformation $\chi=F[\vp]$ in order to compute
$\Gamma_{\vp,k}[\chi]$. This relation,
\bel{T1}
\langle\chi'\rangle=F[\vp]\ ,
\ee
may differ from the microscopic relation
\begin{align}
\label{T2}
\vp'=&\,H[\chi']=N(\chi')\chi'\ ,\nn\\
\chi'=&\,H^{-1}[\vp']=N^{-1}[\vp']\vp'\ .
\end{align}
One can address these issues within the two-field formalism introduced in
sect.~\ref{sec:FTEA}. 

\subsection*{Effective average action for fundamental and\\composite fields}

The two-field effective action in eq.~\eqref{E28} can be extended by adding an
infrared cutoff terms $\cR_{k,\chi}$ for the fundamental fields, and
$\cR_{k,\vp}$ for the composite fields
\begin{align}
\label{T3}
&\Gamma_k'[\chi,\vp]=-\ln\int\cD\chi'\,\exp\bigg\{-S[\chi']\nn\\
&-\frac12\int_{x,y}\Big[\chi'(x)\cR_{k,\chi}(x,y)\chi'(y)\nn\\
&\quad\quad\quad\quad+H\gl\chi'\gr(x)\cR_{k,\vp}(x,y)H\gl\chi'\gr(y)\Big]\nn\\
&+\int_x\bigg[\frac{\partial\Gamma_k'}{\partial\chi(x)}\gl\chi'(x)-\chi(x)\gr\nn\\
&\quad\quad+\frac{\partial\Gamma_k'}{\partial\vp(x)}\gl
H\gl\chi'\gr(x)-\vp(x)\gr\bigg]\bigg\}\ .
\end{align}
For $\cR_{k,\chi}=0$ and $\partial\Gamma_k'/\partial\chi(x)=0$ one recovers the
effective action $\Gamma_{\vp,k}[\vp]$ defined by eq.~\eqref{C1}. We may define
$\Gamma_{\vp,k}[\chi,\vp]$ by setting in eq.~\eqref{T3} the infrared cutoff for
the microscopic fields to zero, $R_{k,\chi}=0$. Solving the field equation for
$\chi[\vp]$,
\bel{T4}
\frac{\partial\Gamma_{\vp,k}'[\chi,\vp]}{\partial\chi(x)}\gl\chi_0[\vp]\gr=0\ ,
\ee
one recovers
\bel{T5}
\Gamma_{\vp,k}'[\vp]=\Gamma_{\vp,k}'\big[\chi_0[\vp],\vp\big]\ .
\ee

On the other hand, additional information contained in
$\Gamma_{\vp,k}'[\chi,\vp]$ for arbitrary $\chi$ permits us to compute
correlation functions for $\chi'$ in the presence of sources for $\chi'$ and
$\vp'$, and in the presence of the IR-cutoff $\cR_{k,\vp}$. These are the
relations
\bel{T7}
\chi(x)=\langle\chi'(x)\rangle\ ,
\ee
or
\bel{T8}
\int_y\langle\chi'(x)\chi'(y)\rangle_c\frac{\partial^2\Gamma_{\vp,k}'[\chi,\vp]}{\partial\chi(y)\partial\chi(z)}=\delta(x-z)\
.
\ee
In particular, the solution of the field equation for $\chi$ in eq.~\eqref{T4}
defines $\langle\chi'\rangle=F[\vp]$ by
\bel{T9}
\langle\chi'\rangle=\chi_0[\vp]=F[\vp]\ ,
\ee
since for this solution the source term for the microscopic field vanishes.

We emphasize that the relation $\langle\chi'\rangle=F[\vp]$ cannot be obtained
from the effective average action $\Gamma_{\vp,k}$. It needs the additional
information provided by the two-field formalism.

\subsection*{Initial value problem}

The two field formalism sheds light on the initial value problem for
$\Gamma_{\vp,k}$. We will see that for the initial value of the flow at large
$k$ or $k\to\infty$ the issue of the Jacobian in the microscopic field
transformation for the functional integral comes back in an indirect way. For
this finding it does not matter if we change the microscopic integration
variables or not. The flow equation for $\Gamma_{\vp,k}[\chi,\vp]$ is the same
as for $\Gamma_{\vp,k}[\vp]$, e.g. eq.~\eqref{C19}, except for the additional
argument $\chi$ of the effective action. For its solution we need the initial
value of $\Gamma_{\vp,k}[\chi,\vp]$ for a large value $k=\Lambda$ or
$k\to\infty$. The initial value of $\Gamma_{\vp,\Lambda}[\vp]$ is then obtained
from eq.~\eqref{T5}.

By subtraction of the macroscopic cutoff term the functional expression for
$\Gamma_{\vp,k}[\chi,\vp]$ reads
\begin{align}
\label{T10}
&\Gamma_{\vp,k}[\chi,\vp]=\Gamma_{\vp,k}'[\chi,\vp]-\frac12\int_{x,y}\vp(x)\cR_{k,\vp}(x,y)\vp(y)\nn\\
=&\,-\ln\int\cD\chi'\,\exp\bigg\{-S[\chi']\nn\\
&-\frac12\int_{x,y}\big[H\gl\chi'\gr(x)-\vp(x)\big]\cR_{k,\vp}(x,y)\big[H\gl\chi'\gr(y)-\vp(y)\big]\nn\\
&+\int_x\bigg[\frac{\partial\Gamma_{\vp,k}}{\partial\vp(x)}\big[H\gl\chi'\gr(x)-\vp(x)\big]\nn\\
&\quad\quad+\frac{\partial\Gamma_{\vp,k}}{\partial\chi(x)}\big[\chi'(x)-\chi(x)\big]\bigg]\bigg\}\
.
\end{align}
For $k\to\infty$ the exponent in the functional integral is dominated by the
term $~\sim\cR_{k,\vp}\sim k^2$. We want to evaluate
$\Gamma_{\vp,\Lambda}[\chi,\vp]$ in the limit $k=\Lambda\to\infty$.

For $k\to\infty$ we employ $H(\chi')=F^{-1}(\chi')$ and define $\hat\chi$ by
\bel{T11}
\chi'=F(\vp)+\hat\chi\ .
\ee
One infers
\bel{T12}
H(\chi')=F^{-1}\gl F(\vp)+\hat\chi\gr=\vp+A(\vp)\hat\chi+\hdots
\ee
where
\bel{T13}
A(\vp)=\frac{\partial H}{\partial\chi'}\gl F(\vp)\gr\ .
\ee
Keeping only the leading terms and making a shift in the integration variable
yields
\bel{T14}
\Gamma_{\vp,k}[\chi,\vp]=S\big[F(\vp)\big]-\int_x\frac{\partial\Gamma_{\vp,k}}{\partial\chi(x)}\gl
F(\vp)-\chi\gr(x)+\Delta\Gamma_k[\chi,\vp]\ ,
\ee
with
\begin{align}
\label{T15}
&\Delta\Gamma_k[\chi,\vp]=-\ln\int\cD\hat\chi\,\exp\bigg\{\nn\\
&\quad\quad-\frac12\int_{x,y}\gl A(\vp)\hat\chi\gr(x)\cR_{k,\vp}(x,y)\gl
A(\vp)\hat\chi\gr(y)\nn\\
&+\int_x\bigg[\frac{\partial\Gamma_{\vp,k}}{\partial\vp(x)}A\gl\vp\gr(x)+\frac{\partial\Gamma_{\vp,k}}{\partial\chi(x)}-\frac{\partial
S[\chi]}{\partial\chi(x)}\gl F(\vp)\gr\bigg]\hat\chi(x)\bigg\}\ .
\end{align}

We focus on solutions of the partial field equation~\eqref{T4} for which one
infers
\bel{177A}
\Gamma_{\vp,k}[\vp]=S\big[F(\vp)\big]+\Delta\Gamma_k\big[\chi_0(\vp),\vp\big]\ .
\ee
In the limit $k\to\infty$ we can take $\cR_k(x,y)=k^2\delta(x-y)$. Contributions
from the term linear in $\hat\chi(x)$ in the exponent in eq.~\eqref{T15} are
suppressed by $k^{-2}$ and we end with the Gaussian functional integral
\bel{T16}
\Delta\Gamma_k[\chi,\vp]=-\ln\int\cD\chat\,\exp\bigg\{-\frac12\int_xk^2A^2\gl\vp\gr(x)\chat^2(x)\bigg\}\
.
\ee
The field-dependent part of this term reads
\begin{align}
\label{T17}
\Delta&\Gamma_k[\chi,\vp]=-\ln\det A^{-1}(\vp)\nn\\
=&\,-\tr\ln A^{-1}(\vp)=\tr\ln A(\vp)=\int_x\eps^{-4}\ln\big[A\gl\vp\gr(x)\big]\
.
\end{align}
It depends on $\chi$, but not on $\vp$. With
\bel{T18}
A(\vp)=\frac{\partial\vp'}{\partial\chi'}\bigg|_{F(\vp)}\ ,
\ee
we realize that a term corresponding to the Jacobian in
eqs.~\eqref{M10}-~\eqref{M13} arises now on the macroscopic level. The initial
value of $\Gamma_{\vp,k}[\vp]$ for $k\to\infty$ involves the Jacobian of the
non-linear microscopic field transformation.

This is consistent with the fact that one could have done a transformation of
integration variables on the microscopic level. With a functional integral over
$\vp'$ and a classical action $\Sbar[\vp']$, the limit $k\to\infty$ of the
effective average action results directly in
$\Gamma_{k\to\infty}[\vp]=\Sbar[\vp]$, which contains the contribution of the
Jacobian in eq.~\eqref{M10}. Unfortunately, one recovers the
dependence on the regularization encoded in $\eps^{-4}$. This may have the
consequence that the effective action $\Gamma_{\vp,k}$ may be difficult to
handle in practice for the extraction of macroscopic information for a given
microscopic model. In particular, starting with $S[\chi']$ involving only a
non-linear kinetic term~\eqref{M1} it seems unavoidable that the microscopic
effective action $\Gamma_{\vp,k\to\infty}$ involves a scalar potential and does
not correspond to a model of a free massless scalar field. In summary, the
initial values of the effective average action $\Gamma_{\vp,k}[\vp]$ are
precisely the ones corresponding to a variable transformation in the functional
integral. As long as the variable transformation $\vp'[\chi']$ is independent of
$k$ one retains the standard flow equation. The correction terms in
eq.~\eqref{C19} arise if one decides to make the microscopic field
transformation $k$-dependent.

\subsection*{Fundamental and composite fields}

We conclude that the effective action and the functional flow equations often
single out in practice a preferred choice of \qq{canonical or fundamental
fields}. These are the fields for which the measure is linear in the fields,
$\int\cD\chi'=\int\cD\chat$, for $\chat=\chi'-\chi$, and for which the classical
action $S[\chi']$ is supposed to be known. These are typically the fields in
terms of which a given model is formulated. In principle, microscopic field
transformations to \qq{composite fields} which depend non-linearly on $\chi'$
are
possible. They require, however, control of a Jacobian, which is typically not
given in practice. This holds in an obvious way for the variable transformations
in the functional integral. The problem reappears if we define the effective
action and flow equations with sources and infrared cutoff for the composite
field. It now concerns the initial value of the flow for large $k$. This
inherits the Jacobian even if we keep the fundamental field $\chi'$ as the
integration variable. The Jacobian can be avoided if for $k=\Lambda$ the
composite field coincides with the fundamental field.

While the two-field formalism finds its limitations if restricted to a vanishing
infrared cutoff for $\chi'$, it can provide important information and
simplifications if we also introduce a non-vanishing cutoff $\cR_{k,\chi}$ for
the fundamental field $\chi'$. Having at our disposal two cutoffs $\cR_{k,\chi}$
and $\cR_{k,\vp}$, we can arrange the setting such that for $k\to\infty$ only
$\cR_{k,\chi}$ matters, implying the simple initial value of the microscopic
effective action
\bel{T22}
\Gamma_{k\to\infty}[\chi,\vp]=S[\chi]\ .
\ee
The infrared cutoff $\cR_{k,\vp}$ may only become relevant at lower $k$ where
composite fields as $\vp$ play an important role. Using suitable identities one
can reformulate the two-field effective action as a functional integral over two
fields $\chi'$ and $\vp'$.~\cite{FW}. Both fields are then formally on equal
footing. This type of approach has found wide applications, for example for the
description of mesons in a model of strong interactions involving quarks and
gluons as fundamental fields~\cite{GW2, ELWE, JUWE1, DJW, JUWE2, MEWE, BRTE,
BGI, BRA, BHMP, HPS, HSBP, MISCH, HMP, FPR}, see also~\cite{FKK, MSFR, EBL,
HPS2, MPS, EGR, STW, RPS, BLP1, BLPAW, AMMM, JAPA, BLP2, FEPA, BRSCH}.

\section{Scaling solutions}\label{sec:SS}

The functional flow equations may admit scaling solutions for which
dimensionless functions of suitable renormalized dimensionless fields do no
longer depend on $k$. In this way a given model can be extrapolated to
arbitrarily large $k$, thereby constituting an ultraviolet completion. The limit
$k\to\infty$ corresponds to an ultraviolet fixed point. The existence of the
ultraviolet fixed point induces renormalizability of the theory. If the
interactions are not small at the fixed point one encounters non-perturbative
renormalizability or asymptotic safety~\cite{WEIN}.

As long as one is interested only in the existence of possible scaling solutions
or fixed points and associated universality classes, no knowledge of the
microscopic effective action $\Gamma_k$ for $k\to\infty$ is needed. The
corresponding form of $\Gamma_{k\to\infty}$ is inferred a posteriori from the
scaling solution. The freedom of non-linear field transformations may be helpful
to find scaling solutions. In particular, these transformations can be very
useful in order to identify the renormalized fields in terms of which the
scaling solution becomes independent of $k$. For the microscopic field
transformations in the flow equation a non-linear RG-kernel $\gamma[\vp]$ can
reduce the complexity of the flow equation and facilitate the detection of
possible scaling solutions. Having found a scaling solution establishes a
universality class for the corresponding model. The existence of the
universality class is often sufficient for many purposes. No detailed
microscopic realization of this universality class may be needed. 

For our simple scalar model the use of suitable RG-kernels for $\Gamma_{\vp,k}$
permits the statement that there exists the universality class of a free
massless scalar field, corresponding to the choice of the microscopic effective
action
\bel{T21}
\Gamma_{\vp,k\to\infty}=\frac12\int_x\partial^\mu\vp\partial_\mu\vp\ .
\ee
This statement is rather trivial. Since the relation between $\chi'$ and $\vp$
is not known this does not help much to answer the question what is the physics
of the model given by eqs.~\eqref{M1},~\eqref{M4}. Eq.~\eqref{T21} is related to
eq.~\eqref{eq:AA} by a macroscopic field transformation and therefore compatible
with the existence of such an ultraviolet fixed point. For the interesting
question if a crossover trajectory links this fixed point to the trivial fixed
point for $k\to0$ an elimination of the kinetial by a suitable RG-kernel does
not seem to be helpful.

For more complex settings the possible scaling solutions and universality
classes may not be known. An example is quantum gravity. In this case the flow
equation~\eqref{C19} for $\Gamma_{\vp,k}$ could be helpful in order to identify
candidates for scaling solutions. The use of an arbitrary functional
$\gamma[\vp]$ in order to eliminate unwanted couplings (sometimes called
\qq{inessential couplings}) is accompanied, however, by two issues. The first
concerns the fact that for arbitrary $\gamma[\vp]$ the relation between
fundamental and composite fields is not known explicitly. In particular, there
is no control if this relation is invertible or not. As a consequence, it is not
guaranteed that a scaling solution found for unrestricted $\gamma[\vp]$ is
indeed a scaling solution for the model. Promoting such a candidate to a full
scaling solution would require to find the microscopic field transformation that
leads to the form of $\gamma[\vp]$ used, and to establish its regularity.

One may argue that it is sufficient to find a solution of the coupled system of
flow equations~\eqref{C18},~\eqref{C19}, without the need of the existence of an
associated microscopic field transformation. Eq.~\eqref{C18} is then taken
purely as a change in the space of macroscopic fields $\vp$, without the
realization that $\vp$ is the expectation value of some microscopic composite
field. The problem with this view is that one could replace in this case the
flow equation~\eqref{C19} by many other forms of modified flow equations. The
specific form of eq.~\eqref{C19} is rooted in the microscopic formulation of the
functional integral~\eqref{C5}. If no such functional integral exists, the
status of $\Gamma_{\vp,k}$ and its relation to correlation functions becomes
unclear. It is no longer guaranteed that some microscopic realization exists for
the ultraviolet completion associated to the scaling solution.

In short, many of the scaling solutions found for the \qq{essential
renormalization group} may be fake, either because no microscopic realization
exists, or due to a shift of important physical information into sectors that
are not resolved sufficiently by the truncation. This seems to happen for
quantum gravity if higher order invariants based on polynomials of the curvature
tensor are absorbed by the RG-kernel of microscopic field transformations. In
this approach a large number of scaling solutions appears formally~\cite{BKN2,
BFK}. It is not a simple issue to decide which ones are physical or fake.

As a second issue the scaling solutions found by microscopic field
transformations with arbitrary non-linear RG-kernels may not be complete. For
our case of a scalar theory one could define $\gamma[\vp]$ such that in first
order of a derivative expansion the effective action is restricted effectively
to a potential and a kinetic term with constant coefficient. The (approximate)
scaling solution for this model in four dimensions is a \qq{trivial theory} for
which the quartic scalar coupling flows slowly towards zero as $k\to0$. Due to a
Landau pole this theory is not \qq{ultraviolet complete} in the sense that it
cannot be extended to $k\to\infty$. The only full scaling solution in this
setting is the free massless scalar field. This statement does not answer the
question if additional scaling solutions for the models of the type
of~\eqref{M1} could exist. If scaling solutions with a non-trivial field
dependence of the kinetial of the type discussed in
appendices~\ref{app:FEK},~\ref{app:FKP} exist, the transformation to a canonical
kinetic term may have moved the relevant information to the part that has been
discarded by the truncation.

An example of missing scaling solutions for the essential renormalization group
is asymptotic freedom in quantum gravity. In the presence of invariants
quadratic in the curvature tensor quantum gravity is
renormalizable~\cite{KS1977} and asymptotically free in perturbation
theory~\cite{FT1982, AB1985}. By use of functional renormalization it has been
possible to follow the flow of the relevant couplings into and inside the
non-perturbative domain~\cite{SWY1}. Scaling solutions have been found which
link
the asymptotically free ultraviolet fixed point to an infrared fixed
point~\cite{SWY2}. For the essential renormalization group the couplings
multiplying the terms quadratic in the curvature tensor are eliminated by a
field transformation. It seems hard to find asymptotic freedom and the
corresponding scaling solution in this approach.

\section{Discussion}\label{sec:C}

The present paper addresses a systematic discussion of field transformations,
starting from variables in the functional integral and proceeding to
transformations of field variables in the effective action and for flow
equations for associated scale-dependent effective average actions. In order to
illustrate the conceptual discussion with a practical case we have focused often
on a simple scalar model with a field-dependent kinetic term. For our concluding
discussion we focus on field transformations for functional flow equations.

Transformations of the field variables for functional flow equations are a
powerful and versatile tool, both for technical simplifications and the
understanding of key physical features. General non-linear field transformations
open, however, a very wide area. Pushed to extremes, especially for cases where
the regularity of the field transformation is not under control, they entail the
danger that important physical information becomes hidden. In case of
approximations to the most general form of the effective action (truncations)
key physical information may even get lost.

For the field transformations in functional flow equations we distinguish
between two different concepts -- the macroscopic and the microscopic field
transformations. The macroscopic field transformations are transformations of
the variables in terms of which a given effective average action is expressed.
They do not change the effective average action and the associated field
equations and macroscopic correlation functions. These quantities are only
expressed in terms of different \qq{field coordinates}. As familiar from
differential equations, a suitable choice of variables can be very helpful to
find (approximate) solutions. In quantum field theory the transformation to
renormalized fields can absorb part of the microscopic information which does
not affect the macroscopic observations. The use of dimensionless renormalized
fields reveals the possible existence of scaling solutions. Selecting suitable
fields can be of great help to make the relevant macroscopic degrees
of freedom easily visible.

The advantage of the use of macroscopic field transformations is the explicit
control of the transformation at every stage. On the other hand, the fact that
the effective average action is not changed may be viewed as a disadvantage.
Macroscopic field transformations do not adapt the cutoff to suitable non-linear
composite fields. For non-linear field transformations the appearance of
connection terms in second functional derivatives may sometimes constitute a
further technical disadvantage.

Microscopic field transformations change the effective average action beyond a
simple change of variables. This change results from the use of different cutoff
functions and different sources employed in the defining functional integral for
the effective average action. For non-linear field transformations a non-trivial
difference between the field-transformed and the original effective action
remains in the limit $k\to0$ for which all fluctuations are integrated out. The
difference in the source terms remains even if the cutoff is removed.

For linear microscopic field transformations the different source terms simply
result in a linear macroscopic field transformation. For linear transformations
the macroscopic and microscopic field transformations are equivalent. This does
no longer hold for non-linear transformations. Macroscopic correlation functions
are defined by functional derivatives of the effective action. In the non-linear
case there is no longer a direct map between the macroscopic correlation
functions and microscopic correlations, neither for the original canonical field
nor for composite microscopic fields. This contact is can be reconstructed for
the macroscopic field transformations as long as the transformation to canonical
fields is known explicitly. Microscopic field transformations of the average
effective action typically loose the direct contact to the microscopic
correlation functions for the canonical fields of the original model. This is
also no direct relation to correlations of microscopic composite fields if the
relation between the canonical and composite fields is not known for a given
RG-kernel.

Microscopic field transformations also affect the initial conditions for the
flow at $k\to\infty$ or very large $k$. Relating this \qq{microscopic effective
average action}, e.g. $\Gamma_k$ for very large $k$, to the \qq{classical
action} $S$ used to define the functional integral for the model involves a
Jacobian. This Jacobian is the same as the one appearing for a transformation of
the field variables in the functional integral. The Jacobian remains
field-independent only for linear field transformations. It can therefore be
omitted only if at some \qq{initial} or \qq{microscopic} scale $\Lambda$ the
composite field equals the fundamental field modulo a linear field
transformation.

A simple possible view on the microscopic field transformations associates them
to transformations of the integration variables in the functional integral. As
long as variable transformations do not depend on $k$ the flow equations are not
affected. The correction terms involving the RG-kernel $\gamma[\vp]$ arise from
the $k$-dependence of the field transformations. The additional terms can be
associated to an additional $k$-dependence resulting from a $k$-dependent
Jacobian, combined with a $k$-dependent macroscopic field transformation. The
difference of the modified flow equations for non-linear macroscopic and
microscopic field transformations results from the Jacobian which depends both
on fields and on $k$. This explains why for linear field transformations the
macroscopic and microscopic transformations are identical for suitably chosen
cutoff functions.

The advantage of the use of microscopic field transformations is the great
flexibility if arbitrary non-linear RG-kernels are used. In particular, the
cutoff can be adapted to composite fields reflecting the relevant degrees of
freedom. Important technical improvements can be achieved by employing the
correction terms in order to remove technically annoying terms from the flow
equations. The disadvantage of the microscopic field transformations with
non-linear RG-kernel is the loss of control of the transformation. It is
generally not known which microscopic field transformation produces a given
RG-kernel. This relation can be rather complicated. Regular microscopic field
transformations impose constraints on the allowed RG-kernels. These constraints
are not known in practice, even not approximately, such that solutions of the
flow equations obtained for arbitrary $\gamma[\vp]$ are not guaranteed to be
valid.

The lack of direct access to the microscopic field transformation makes it hard
to follow to which sector of the effective average action a given physical
information is moved by the use of arbitrary non-linear $\gamma[\vp]$. This can
be rather dangerous for truncated solutions since important information can be
moved to the part which is truncated away.

Finally, the initial values for the flow equation encoded in the microscopic
effective average action are connected to the classical action defining the
functional integral of the model by a relation involving a Jacobian that may be
difficult to handle. This disadvantage does not affect the search for scaling
solutions and associated universality classes.

For the search for scaling solutions the removal of technical complications by
the use of suitable RG-kernels can enlarge the accessible space of truncations.
This may contribute to the robustness of scaling solutions found. On the other
hand, some of the scaling solutions can be discarded by the truncation and may
no longer be visible if arbitrary RG-kernels are employed. The set of scaling
solutions found in this way may not be complete. The set of scaling solutions
found for the flow equations of the \qq{essential renormalization group} may
also
contain fake solutions which do not correspond to scaling solutions of the
model. Obstructions may be overlooked if the regularity constraints on
$\gamma[\vp]$ are discarded. While the set of all scaling solutions for the
essential renormalization group may be neither all valid nor complete, there
remains a good chance that relevant scaling solutions are found.

The view that couplings which can be removed by the use of suitable RG-kernels
are not observable is not justified. Field transformations only move physical
information between different sectors. The expression of a given observable
in terms of fields and functional derivatives of the effective action is
affected by a field transformation.

Both the macroscopic and the microscopic field transformations have their
advantages and disadvantages, justifying their use adapted to a given purpose. A
setting that combines the advantages and removes disadvantages is the two-field
formalism using both fundamental and composite fields. We have not discussed
here this formalism in detail and refer to ref.~\cite{FW} (or ref.~\cite{PAW}
for similar considerations). The prize to pay for the two-field formalism is the
higher complexity due to additional couplings for a description in terms of two
fields.

The present paper partly answers the question if the functional flow equation
distinguishes between fundamental and composite fields. A first observation
states that the correction terms arise only from the $k$-dependence of field
transformations. Fields that are related by a $k$-independent field
transformation induce identical flow equations. On this level the functional
flow equation does not single out a particular choice of fields. We may call
fields \qq{canonical} if the classical action is expressed in terms of the
canonical fields $\chi'(x)$ and the functional measure is linear in $\chi'(x)$,
$\cD\big[\chi'(x)+a(x)\big]=\cD\chi'(x)$. Fields related to $\chi'(x)$ by a
non-linear field transformation are no longer canonical. For many purposes one
may associate the canonical fields with fundamental fields and call fields
obtained by a non-linear field transformation composite. This naming is
appropriate if the fundamental fields are the fields in terms of which a model
is formulated. We note, however, that there is no fundamental meaning to this
concept since for a different choice of the effective action the composite
fields may become canonical. On a conceptual level all choices of fields are
equivalent.

On the conceptual level there is no difference between canonical and
non-canonical fields. Switching to the classical action $\Sbar[\vp']$ and
defining the effective action as $\Gamma_\vp[\vp]$ the non-canonical field
becomes canonical. In practice, a given canonical field is often associated with
the fundamental field. The non-canonical field is then considered as a composite
field. A $k$-dependent field transformation is no longer defined purely on the
microscopic level. Fields related to the canonical fundamental field by a
$k$-dependent field transformation can always be called \qq{composite fields} or
\qq{dressed fields}. In this case the appearance of correction terms in the flow
equation indicates \qq{compositeness} of the fields. These dressed fields
include
the renormalized fields. We infer from this short discussion that the notion of
composite fields depends on the context.

We have discussed scalar field theories with a field-dependent kinetial $K$.
This serves as a concrete example for the role of non-linear field
transformations on different levels and for different versions. In particular we
propose that for small values of the scalar field $\chi$ there exist scaling
solutions where $K\sim\chi^{-2}$. They are connected to an ultraviolet fixed
point with the quantum scale symmetry of a multiplicative rescaling of $\chi$.
If this fixed point is confirmed, and if a flow trajectory relates it to the
trivial fixed point for $\chi\to\infty$ or $k\to0$, this setting can lead to an
ultraviolet complete scalar quantum field theory with interactions in four
dimensions, overcoming the triviality problem.

The most direct evidence for this fixed point comes from the direct computations
in app.~\ref{app:FEK},~\ref{app:FKP} with a field-dependent cutoff respecting
the
scale symmetry. One may ask if non-linear field transformations can yield
additional evidence for the existence of this fixed point. A simple part of the
answer is the statement that there exist versions of the effective action and
the flow equations for which this fixed point can be firmly established. This is
rather trivial, since for these versions one deals with a free theory for
composite scalar fields. The interesting question asks if a free theory for
composite fields in the ultraviolet can be connected by a flow trajectory to a
free theory for fundamental fields in the infrared. For this issue the field
transformations have so far not brought substantial additional insight.
Nevertheless, the present paper sets the stage for an investigation of this
question.

Despite our sometimes critical assessment of the dangers of non-linear field
transformations we consider such field transformations as a very rich and
powerful tool for the understanding of the physics of a given model. Care is
needed, however, in order to keep limitations in mind and to refrain from too
general statements.

\subsection*{Acknowledgment}

The author would like to thank J. Pawlowski for fruitful discussions.

\appendix

\section{Flow equation for field dependent\\kinetial}\label{app:FEK}

In this appendix we discuss the flow equation for a field-dependent coefficient
of the kinetic term or \qq{kinetial} $K$ in the \qq{leading kinetial
approximation}. For this truncation only the term with two derivatives is kept
in the effective average action, while the potential and higher derivative terms
are omitted. This truncation could be valid in regions of field space for which
$K$ is very large and depends on $\chi$. Renormalized couplings for
non-derivative interactions scale with inverse powers of $K$ and could therefore
only have a small effect. A field-independent $K$ can be absorbed as a wave
function renormalization, but interactions due to the $\chi$-dependence of the
kinetial do matter. One may check the validity of the leading kinetial
approximation by computing the induced potential and investigate its role on the
flow of the kinetial $K$. This will be done in the appendix~\eqref{app:FKP}.

In the present appendix no field transformations are performed. This direct
computation of the flow of the kinetial can serve as a point of comparison for
computations using field transformations. We briefly discuss possible scaling
solutions in the leading kinetial approximation. In this truncation we find
evidence for the ultraviolet fixed point~\eqref{eq:AA}. We treat here $\chi$ as
the canonical scalar field for which the flow equation~\eqref{F8} holds without
corrections. As long as field transformations remain independent of $k$ the same
flow equation holds for composite fields as $\vp\sim\ln\chi$. For
$k$-independent flow equations the microscopic Jacobian does not influence the
flow equation. It matters only for the initial conditions for the solutions for
$k\to\infty$.

\subsection*{Flow equation}

The flow equation for the truncation~\eqref{F31} can be inferred from the
computation in ref.~\cite{CWDE, SEWE, GEW} by omitting contributions from the
potential. We focus on four dimensions, $d=4$, where one finds
\begin{align}
\label{FG1}
\partial_tK(\chi)=&\,-\frac{k^2}{32\pi^2K_0^2}\left(\frac{\partial
K}{\partial\chi}\right)^2\nn\\
&\times\left[m_4^8\left(0,\eta_K,\frac{K}{K_0}\right)-\frac92l_2^6\left(0,\eta_K,\frac{K}{K_0}\right)\right]\nn\\
&-\frac{k^2}{16\pi^2K_0}\frac{\partial^2K}{\partial\chi^2}l_1^4\left(0,\eta_K,\frac{K}{K_0}\right)\
.
\end{align}
Here $K_0$ is a type of reference kinetial that will be defined later. It enters
the definition of the cutoff function
\bel{FG2}
\cR_k(q^2)=K_0k^2r(y)\ ,\quad y=\frac{q^2}{k^2}=\frac{x}{k^2}\ ,
\ee
with \qq{cutoff anomalous dimension} $\eta_K$ defined by
\bel{FG3}
\eta_K=-\partial_t\ln K_0\ .
\ee

The momentum integrals $l_1^4$, $l_2^6$ and $m_4^8$ are defined as
\begin{align}
\label{FG4}
l_1^4\left(0,\eta_K,\frac{K}{K_0}\right)=&\,\frac{K_0}{2k^2}\int_0^\infty\text{d}x\,x\frac{\partial_t\cR_k(x)}{P^2(x)}\
,\nn\\
l_2^6\left(0,\eta_K,\frac{K}{K_0}\right)=&\,\frac{K_0^2}{k^2}\int_0^\infty\text{d}x\,x^2\frac{\partial_t\cR_k(x)}{P^3(x)}\
,\nn\\
m_4^8\left(0,\eta_K,\frac{K}{K_0}\right)=&\,\frac{2K_0^2}{k^2}\int_0^\infty\text{d}x\,x^4\bigg[\frac{\partial_t\cR_k(x)}{P^{5}(x)}\gl\partial_xP(x)\gr^2\nn\\
&-\frac1{2P^4(x)}\partial_xP(x)\partial_t\partial_x\cR_k\bigg]\ .
\end{align}
They involve the inverse propagator in the presence of the cutoff
\bel{FG5}
P(x)=Kx+\cR_k=K_0k^2\gl\tilde zy+r(y)\gr\ .
\ee
Here we employ for the \qq{normalized kinetial} the ratio
\bel{FG6}
\tilde z=\frac{K}{K_0}\ .
\ee

\subsection*{Threshold functions}

The momentum integrals (often called threshold functions) carry key information
of the flow equation. Using
\bel{FG6}
\partial_t\cR_k=K_0k^2\left(2-\eta-2\frac{\partial\ln r}{\partial\ln y}\right)r
\ee
the factors $K_0^p$ cancel such that for a given choice of $r(y)$ these
dimensionless integrals depend only on $\eta$ and $\tilde z$. (We omit the index
on $\eta$ from now on.) The powers of $k$ drop out if we switch to dimensionless
integration variables $y=x/k^2$,
\begin{align}
\label{FG7}
l_1^4(0,\eta,\tilde z)=&\,\frac12\int_0^\infty\text{d}y\,y(\tilde
zy+r)^{-2}\big[(2-\eta)r-2y\partial_yr\big]\ ,\nn\\
l_2^6(0,\eta,\tilde z)=&\,\int_0^\infty\text{d}y\,y^2(\tilde
zy+r)^{-3}\big[(2-\eta)r-2y\partial_yr\big]\nn\\
=&\,-\frac{\partial}{\partial\tilde z}l_1^4(0,\eta,\tilde z)\ ,\nn\\
m_4^8(0,\eta,\tilde z)=&\,2\int_0^\infty\text{d}y\,y^4(\tilde zy+r)^{-5}(\tilde
z+\partial_yr)^2\nn\\
&\qquad\times\big[(2-\eta)r-2y\partial_yr\big]\nn\\
+2\int_0^\infty\text{d}&y\,y^5(\tilde zy+r)^{-4}(\tilde
z+\partial_yr)\gl\partial_y^2r+\frac{\eta}{2y}\partial_yr\gr\ .
\end{align}

As an example, we choose the Litim cutoff function~\cite{LIT}
\bel{FG8}
r(y)=(1-y)\theta(1-y)\ ,
\ee
for which ($\tilde z\neq0$)
\bel{FG9}
l_1^4(0,\eta,\tilde
z)=\frac12\int_0^1\text{d}y\,y\gl1+(\ztil-1)y\gr^{-2}\gl2-\eta(1-y)\gr
\ee
The Litim cutoff needs to be defined as the limiting case of smooth functions in
order to make $\partial_y^2r\partial_yr$ well defined. For our choice one finds
\begin{align}
\label{FG10}
m_4^8(0,\eta,\tilde z)=&\,-\frac16(\ztil-1)^2\frac{\partial^3l_1^4(0,\eta,\tilde
z)}{\partial\ztil^3}+\frac2{\ztil^3}-\frac1{2\ztil^4}\nn\\
&\,-\eta(\ztil-1)\int_0^1\text{d}y\,y^4\big[1+(\tilde z-1)y\big]^{-4}\ .
\end{align}
Evaluating the integral yields
\begin{align}
\label{FG11}
l_1^4(0,\eta,\tilde
z)=&\,\frac{\ln\ztil}{(\ztil-1)^2}-\frac1{\ztil(\ztil-1)}\nn\\
&-\frac\eta{(\ztil-1)^2}\left(\frac12\ln\ztil-1+\frac1{\ztil-1}\ln\ztil\right)\
.
\end{align}
For $\ztil=1$ the integrals simplify
\begin{align}
\label{FG12}
\hat l_1^4(\eta)=&\,l_1^4(0,\eta,\ztil=1)=\frac12-\frac\eta{12}\ ,\nn\\
\hat l_2^6(\eta)=&\,l_2^6(0,\eta,\ztil=1)=\frac23-\frac\eta{12}\ ,\nn\\
\hat m_4^8(\eta)=&\,m_4^8(0,\eta,\ztil=1)=\frac32\ .
\end{align}

For $\eta=0$ one obtains the explicit expressions
\begin{align}
\label{FG13}
\bar l_1^4(\tilde
z)=&\,l_1^4(0,0,\ztil)=\frac{\ln\ztil}{(\ztil-1)^2}-\frac1{\ztil(\ztil-1)}\ ,\\
\bar
l_2^6(\ztil)=&\,l_2^6(0,0,\ztil)=\frac{2\ln\ztil}{(\ztil-1)^3}-
\frac{3\ztil-1}{\ztil^2(\ztil-1)^2}\ ,\nn\\
\bar m_4^8(\ztil)=&\,m_4^8(0,0,\ztil)=\frac{4\ln\ztil}{(\ztil-1)^3}-
\frac{38\ztil^3-19\ztil^2+8\ztil-3}{6\ztil^4(\ztil-1)^2}\ .\nn
\end{align}
For gaining some intuition on the dependence on $\eta$ we also provide the
integrals for $\eta=2$,
\begin{align}
\label{FG14}
l_1^4(0,2,\ztil)=&\,\frac{\ztil+1}{\ztil(\ztil-1)^2}-\frac{2\ln\ztil}{(\ztil-1)^3}\
,\nn\\
l_2^6(0,2,\ztil)=&\,\ztil^{-2}(\ztil-1)^{-3}\left\{2\ztil^2+5\ztil-1-\frac{6\ztil^2}{\ztil-1}\ln\ztil\right\}\
.
\end{align}
There are no poles in these integrals for $\ztil=1$, with a Taylor expansion
around $\ztil=1$ yielding the leading values~\eqref{FG12}.

For large $\ztil$ the effect of the IR-cutoff is suppressed by powers of
$\ztil^{-1}$. In this range of $\ztil$ we can scale out the dominant
$\ztil$-factors by changing the integration variable to $\tilde y=\ztil y$,
\begin{align}
\label{FG15}
l_1^4(0,\eta,\ztil)=&\,\ztil^{-2}\tilde l_1^4(\ztil,\eta)\ ,\nn\\
l_2^6(0,\eta,\ztil)=&\,\ztil^{-3}\tilde l_2^6(\ztil,\eta)\ ,\nn\\
m_4^8(0,\eta,\ztil)=&\,\ztil^{-3}\tilde m_4^8(\ztil,\eta)\ .
\end{align}
Here $\tilde l_1^4(\ztil,\eta)$ obtains from $l_1^4(0,\eta,\ztil)$ by setting in
eq.~\eqref{FG7} $\ztil\to1$, $y\to\tilde y$, $r(y)\to\tilde r(\tilde y)$, and
similar for the other functions. The $\tilde z$-dependence of $\tilde
l_1^4(\ztil,\eta)$ arises only through the argument $\tilde r(\tilde y)=r\gl
y(\tilde y)\gr=r(\tilde y/\ztil)$. Comparison with the full result~\eqref{FG9}
shows a logarithmic dependence of $\tilde l_1^4$ on $\tilde z$. For $\tilde
l_2^6$ one has the relation
\bel{FG16}
\tilde l_2^6(\ztil,\eta)=\left(2-\frac{\partial}{\partial\ln\ztil}\right)\tilde
l_1^4(\ztil,\eta)\ .
\ee
The suppression of the threshold functions slows down the flow. This effect of
\qq{kinetic decoupling} arises since for $\tilde z$ increasing to large values
the effect of the infrared cutoff becomes less and less important.

\subsection*{Field dependent cutoff}

We next discuss a few characteristic solutions of the flow equation for the
kinetial. Besides the dependence on the choice of the shape of the cutoff
function $r(y)$ they also depend on the choice of $K_0$. We discuss here both a
field-dependent and a field-independent choice of $K_0$.

Let us first consider the choice $K_0=K(\chi)$. This is often used effectively
in the \qq{background formalism}. Using a cutoff function depending on the
macroscopic fields induces correction terms in the flow
equation~\cite{wetterich2018gauge, Wetterich_2021, CWSFE}. These correction
terms can be computed in the setting of the simplified flow
equation~\cite{CWSFE}. They have been found to be small and consistent with the
relevant symmetries. We omit these correction terms here. The choice $K_0=K$
brings important technical simplifications since we can use $\ztil=1$ in the
threshold functions. We write eq.~\eqref{FG1} in the form
\begin{align}
\label{FG17}
\partial_tK=&\,-\frac{k^2}{32\pi^2}\bigg\{2\frac{\partial^2\ln
K}{\partial\chi^2}\hat l_1^4(\eta)\nn\\
&+\left(\frac{\partial\ln K}{\partial\chi}\right)^2\left[2\hat l_1^4(\eta)+\hat
m_4^8(\eta)-\frac92\hat l_2^6(\eta)\right]\bigg\}\ ,
\end{align}
with $\hat l_1^4$ etc. given for the Litim cutoff by eq.~\eqref{FG10}. For
$K_0=K$ and $K$ a power of $\chi$ the r.h.s. of eq.~\eqref{FG17} or~\eqref{FG1}
scales $\sim\beta^{-2}$ for $\chi\to\beta\chi$.

For $\eta$ independent of $\chi$ we can make the ansatz of a powerlaw
\bel{FG18}
K(\chi)=c(k)\chi^{-\alpha}\ ,
\ee
for which eq.~\eqref{FG17} becomes
\bel{FG19}
\partial_tc\,\chi^{-\alpha}=-\frac{k^2}{32\pi^2}\chi^{-2}\left[2\alpha\hat
l_1^4+\alpha^2\left(2\hat l_1^4+\hat m_4^8-\frac92\hat l_2^6\right)\right]\ .
\ee
For $\eta$ independent of $\chi$ the $\chi$-dependence drops out for $\alpha=0$
and $\alpha=2$. For $\alpha=0$ we simply have a free field theory with $K$
independent of $k$ or $\chi$. For $\alpha=2$ one finds
\bel{FG20}
\partial_tc=\frac{k^2}{32\pi^2}D(\eta)\ ,\quad \eta=-\partial_t\ln c\ ,
\ee
with $D(\eta)$ given for the Litim cutoff by
\bel{FG21}
D(\eta)=-\gl12\hat l_1^4+4\hat m_4^8-18\hat l_2^6\gr=-\frac\eta2\ .
\ee
Eq.~\eqref{FG20} has two solutions. For the first, $\partial_tc=0$, $\eta=0$ we
find the fixed point solution~\eqref{eq:AA}. This is in agreement with the
vanishing of the loop correction $H=0$ in the saddlepoint
approximation~\eqref{M30},~\eqref{M31}. The second solution
\bel{A23A}
c=\frac{k^2}{64\pi^2}\ ,\quad \eta=-2\ ,
\ee
results in a scaling form
\bel{FG26}
\Gamma_k=\frac{k^2}{128\pi^2}\int_x\frac1{\chi^2}\partial^\mu\chi\partial_\mu
\chi\ .
\ee
Both solutions are invariant under global dilatation or scale transformations
$\chi\to\beta\chi$. For these solutions the kinetial diverges for $\chi\to0$.
The region of small $\chi$ is a candidate for the validity of the leading
kinetial approximation. We will discuss this in appendix~\ref{app:FKP}.

\subsection*{Field independent cutoff}

The solutions of the flow equation become more complex if we take $K_0=K_0(k)$
to be independent of $\chi$. In this case the infrared cutoff term does not
involve the macroscopic field $\chi$ and there are no corrections to the flow
equation. On the other hand, $\ztil=K(\chi)/K_0$ becomes now a field- and
$k$-dependent function. A cutoff with constant $K_0$ is no longer invariant
under dilatations, such that quantum scale invariant solutions of the
type~\eqref{FG19} are at best an approximation. Of course, the free field
solution $K=K_0$ continues to exist. For $\chi$-independent $K_0$ we translate
the flow equation~\eqref{FG1} into a flow equation for $\ztil$,
\begin{align}
\label{FG30}
\partial_t\ztil=&\,\eta\ztil-\frac{k^2}{32\pi^2
K_0}\bigg\{2\frac{\partial^2\ztil}{\partial\chi^2}l_1^4(0,\eta,\ztil)\nn\\
&+\left(\frac{\partial\ztil}{\partial\chi}\right)^2\left(m_4^8(0,\eta,\ztil)
-\frac92l_2^6(0,\eta,\ztil)\right)\bigg\}\ .
\end{align}
We can set $K_0$ to an arbitrary constant by a multiplicative rescaling of
$\chi$.

We first ask if there are \qq{static solutions} with $k$-independent $K(\chi)$
and $K_0$. This requires the r.h.s. of eq.~\eqref{FG1} to vanish, or
\bel{SO1}
2\frac{\partial^2\ztil}{\partial\chi^2}\bar
l_1^4(\ztil)+\left(\frac{\partial\ztil}{\partial\chi}\right)^2\left(\bar
m_4^8(\ztil)-\frac92\bar l_2^6(\ztil)\right)=0\ .
\ee
For the Litim cutoff we employ eq.~\eqref{FG13} in order to obtain an explicit
non-linear second order differential equation for $\ztil(\chi)$,
\begin{align}
\label{SO2}
&2\left(\frac{\ln\ztil}{(\ztil-1)^2}-\frac{1}{\ztil(\ztil-1)}
\right)\frac{\partial^2\ztil}{\partial\chi^2}\nn\\
=&\,\left[\frac{5\ln\ztil}{(\ztil-1)^3}-
\frac{13\ztil^2-5\ztil+3}{6\ztil^4(\ztil-1)}\right]
\left(\frac{\partial\ztil}{\partial\chi}\right)^2\ .
\end{align}
In particular, for $\ztil=1$ one finds
\bel{SO3}
\frac{\partial^2\ztil}{\partial\chi^2}(\ztil=1)=
\frac32\left(\frac{\partial\ztil}{\partial\chi}(\ztil=1)\right)^2\ ,
\ee
with approximate solution near $\ztil=1$ given for $\chi$ near $\chi_1$ by
\bel{SO4}
\ztil(\chi)\approx1-\frac23\ln\left(\frac\chi{\chi_1}\right)\ ,\quad
\ztil(\chi_1)=1\ ,
\ee
or, alternatively, simply $\ztil=1$.

For $\ztil\to\infty$ the leading logarithmic term cancels for
\bel{SO5}
\frac{\partial^2\ztil}{\partial\chi^2}=
\frac{5}{2\ztil}\left(\frac{\partial\ztil}{\partial\chi}\right)^2\ ,
\ee
with approximate solution
\bel{SO6}
\ztil(\chi)\approx\left(\frac{\chi_0}{\chi}\right)^{\frac23}\ .
\ee
This limit should describe the limit of the static solution for $\chi\to0$.

We conclude that the static solutions for a field-independent IR-cutoff differ
substantially from the ones for a field-dependent cutoff with $K_0=K$, where
$K\sim\chi^{-2}$ according to eq.~\eqref{eq:AA}. The origin of this difference
arises from the fact that the field-independent cutoff is \qq{inhomogeneous} in
the sense that the relative weight as compared to $\Gamma_k^{(2)}$ depends
strongly on the value of $\ztil$. For large $\ztil$ the cutoff becomes
subdominant in the combined inverse propagator $P(x)$, rendering it relatively
ineffective. In contrast, for $\ztil\ll1$ the IR-cutoff dominates $P(x)$ for
momenta $q^2\approx k^2$. Furthermore, only the field-dependent cutoff respects
the scale symmetry $\chi\to\beta\chi$ which is a crucial property of the fixed
point~\eqref{eq:AA}. For an investigation of this possible fixed point the
field-dependent cutoff seems clearly more appropriate.

\subsection*{Scaling solution for scale invariant fields}

One is often interested in scaling solutions for which $K$ becomes a function of
the \qq{scaling field} or \qq{scale invariant field}
\bel{SO8}
\tilde\chi=\frac{\chi}{k}\ ,
\ee
without any further $k$-dependence. For a field-dependent cutoff $K_0=K$ the
solution~\eqref{FG26} is of this type. In this case the effective action can be
written in terms of scale invariant fields if one also uses a scale invariant
metric $\tilde g_{\mu\nu}=k^2g_{\mu\nu}$. In contrast, the effective
action~\eqref{eq:AA} only involves the scale invariant field $\tilde\chi$ if
$\tilde g_{\mu\nu}=g_{\mu\nu}$.

We also may investigate scale invariant solutions for a field-independent cutoff
with constant $K_0$. The flow equation for $\ztil$ at fixed
$\tilde\chi$ reads
\bel{SO9}
\partial_t\ztil|_{\tilde\chi}=\partial_t\ztil|_\chi-\frac{\partial\ztil}{\partial\tilde\chi}\partial_t\tilde\chi|_\chi=\partial_t\ztil|_\chi+\tilde\chi\frac{\partial\ztil}{\partial\tilde\chi}\
.
\ee
With primes denoting derivatives with respect to $\tilde\chi$ this yields
\bel{SO10}
\partial_t\ztil=\eta\ztil+\tilde\chi\ztil'-\frac1{32\pi^2K_0}\left\{2\ztil''l_1^4+\gl\ztil'\gr^2\left(m_4^8-\frac92l_2^6\right)\right\}\
.
\ee
(By multiplicative rescaling of $\tilde\chi$ one may set $32\pi^2K_0=2$.) There
is no longer any explicit dependence of the flow equation on the renormalization
scale $k$. The effective action becomes
\bel{SO11}
\Gamma_k[\tilde\chi]=\frac{k^2K_0}{2}\int_x\ztil(\tilde\chi)\partial^\mu\tilde\chi\partial_\mu\tilde\chi\
.
\ee

For a scaling solution $K_0$ is constant, $\eta=0$, $\partial_t\ztil=0$, and
\bel{SO12}
\tilde\chi\ztil'=\frac1{32\pi^2K_0}\left\{2\ztil''\bar
l_1^4(\ztil)+\gl\ztil'\gr^2\left(\bar m_4^8(\ztil)-\frac92\bar
l_2^6(\ztil)\right)\right\}\ .
\ee
With eq.~\eqref{FG13} this reads
\begin{align}
\label{SO13}
32\pi^2K_0\tilde\chi\ztil'=&\,-\gl\ztil'\gr^2\bigg\{5\frac{\ln\ztil}{(\ztil-1)^3}
-\frac{13\ztil^2-5\ztil+3}{6\ztil^4(\ztil-1)}\bigg\}\nn\\
&+\ztil''\left\{\frac{2\ln\ztil}{(\ztil-1)^2}-\frac{2}{\ztil(\ztil-1)}\right\}\
.
\end{align}
Neglecting the l.h.s. for $\tilde\chi\to0$ one finds in this limit the constant
solution $\ztil=\text{const.}$, or the behavior
\bel{SO14}
\ztil=a\tilde\chi^{-\frac23}\ .
\ee
Again, the degree of divergence of the scaling solution for $\tilde\chi\to0$
differs from the behavior $\sim\tilde\chi^{-2}$ found in eq.~\eqref{FG26} for
the field dependent cutoff, reflecting the shortcoming of the inhomogeneity of
the cutoff.

The region of small fields $\tilde\chi\to0$ for solutions with diverging
kinetial could be a candidate for the validity of the leading kinetial
approximation. As for any truncation, one may investigate its validity by
computing the flow of parts in the effective action that have been neglected,
and assess their influence on the flow of the terms retained. In our case, this
concerns in particular the neglected scalar potential which will be included in
appendix~\ref{app:FKP}.

\section{Flow of kinetial and potential}\label{app:FKP}

In this appendix we investigate the functional flow for a single scalar field in
leading order in the derivative expansion. This truncation includes two
field-dependent functions, namely the potential and kinetial. This approximation
has been studied in refs.~\cite{SEWE, GEW} and employed frequently afterwards.
What is new in this appendix is the focus on possible field regions for which
the kinetial diverges. This divergence is not per se a shortcoming since it may
be removed by a different choice of the field variable. In particular, we are
interested in possible scaling solutions which could correspond to an
ultraviolet fixed point defining an ultraviolet complete non-trivial theory.

In first order in a derivative expansion we need to determine the two
dimensionless functions $u'(\tilde\chi)$ and $z(\tilde\chi)$. Here
$u(\tilde\chi)=U(\tilde\chi)/k^4$ is the dimensionless scalar potential and
primes denote derivatives with respect to $\tilde\chi=\chi/k$. In this appendix
we omit the tilde on $z\equiv\tilde z$.

We will find that the leading kinetial approximation discussed in
appendix~\ref{app:FEK} is consistent for a field-dependent cutoff $K_0=K$. In
contrast, the leading kinetial approximation fails for a field-independent
cutoff. This could be taken as an additional point in favor of a field-dependent
cutoff.

\subsection*{Field independent cutoff}

The flow equation for the derivative of the potential at fixed $\tilde\chi$
reads for $K_0=1$, $\eta=0$
\bel{V1}
\partial_tu'=-3u'+\tilde\chi
u''-\frac1{16\pi^2}\left\{z'l_1^6+u'''l_1^4\right\}\ .
\ee
Similarly, the evolution of $z$ is given by
\begin{align}
\label{V2}
\partial_tz=&\,\tilde\chi
z'-\frac1{32\pi^2}\bigg\{2z''l_1^4+(z')^2\bigg(m_4^8-\frac92l_2^6\bigg)\nn\\
&+(u''')^2m_4^4+u'''z'\gl2m_4^6-4l_2^4\gr\bigg\}\ .
\end{align}
The additional terms from the scalar potential involve the third derivative of
$u$.

The threshold functions are modified in the presence of a potential. They are
defined for $n\geq1$ by
\begin{align}
\label{V3}
l_n^d\equiv&\,l_n^d(u'',z)\nn\\
=&\,n\int_0^\infty\text{d}y\,y^{\frac d2-1}\gl zy+r+u''\gr^{-(n+1)}\gl
r-y\partial_yr\gr\ ,
\end{align}
and
\begin{align}
\label{V4}
m_n^d\equiv&\, m_n^d(u'',z)=n\int_0^\infty\text{d}y\,y^{\frac d2}\gl
zy+r+u''\gr^{-(n+1)}\nn\\
&\qquad\qquad\qquad\times\gl z+\partial_yr\gr^2\gl r-y\partial_yr\gr\nn\\
&+2\int_0^\infty\text{d}y\,y^{\frac d2+1}\gl zy+r+u''\gr^{-n}\gl
z+\partial_yr\gr\partial_y^2r\ .
\end{align}
The functions $l_n^d$ obey the recursive relations
\begin{align}
\label{V5}
l_{n+1}^d(w,z)=&\,-(n+\delta_{n,0})^{-1}\frac{\partial}{\partial w}l_n^d(w,z)\
,\nn\\
l_{n+1}^{d+2}(w,z)=&\,-(n+\delta_{n,0})^{-1}\frac{\partial}{\partial
z}l_n^d(w,z)\ ,
\end{align}
with
\bel{V6}
l_0^d(w,z)=\int\text{d}y\,y^{\frac d2-1}\gl zy+r+u''\gr^{-1}\gl
r-y\partial_yr\gr\ .
\ee
We observe the universal property
\bel{V7}
l_n^{2n}(0,z)=z^{-n}\ .
\ee
Similarly, one has for $n\geq1$
\bel{V8}
m_{n+1}^d(w,z)=-\frac1n\frac{\partial}{\partial w}m_n^d(w,z)\ .
\ee
The threshold functions depend on the choice of the infrared cutoff encoded in
$r(y)$.

For the Litim cutoff~\cite{LIT},
\bel{V9}
r(y)=(1-y)\theta(1-y)\ ,
\ee
one finds
\bel{V10}
l_0^d(w,z)=\int_0^1\text{d}y\,y^{\frac d2-1}\big[(z-1)y+1+w\big]^{-1}\ ,
\ee
or
\bel{V11}
l_0^4(w,z)=\frac1{z-1}\bigg\{1-\frac{1+w}{z-1}\ln\bigg(1+\frac{z-1}{1+w}\bigg)\bigg\}\
.
\ee
Expanding in small $(z-1)/(1+w)$ yields
\bel{V12}
l_0^4(w,z)=\frac1{2(1+w)}-\frac{z-1}{3(1+w)^2}+\dots
\ee
The threshold functions $l_n^d$ involved in eqs.~\eqref{V1},~\eqref{V2} follow
from $l_0^4$ by use of eq.~\eqref{V5}. For the Litim cutoff one finds
\begin{align}
\label{V13}
m_4^d(w,z)=&\,4(z-1)^2\int_0^1\text{d}y\,y^{\frac
d2}\big[(z-1)y+1+w\big]^{-5}\nn\\
&+\frac{2z-\frac12}{(z+w)^4}\nn\\
=&\,(z-1)^2l_4^{d+2}(w,z)+\frac{2z-\frac12}{(z+w)^4}\ .
\end{align}
A simple solution of the coupled system of differential
equations~\eqref{V1},~\eqref{V2} is a free massless scalar field with constant
$z$,
\bel{V13A}
u'=0\ ,\quad z=z_0\ .
\ee

\subsection*{Kinetic decoupling}

For large $z\gg1+w$ the threshold functions vanish with inverse powers of $z$.
In the limit of this \qq{kinetic decoupling} we evaluate
\begin{align}
\label{V14}
l_1^4=&\,\frac{\ln z-1}{z^2}\ ,\quad l_2^4=l_1^6=\frac1{z^2}\ ,\quad
l_2^6=\frac{2\ln z-3}{z^3}\ ,\nn\\
m_4^8=&\,\frac{12\ln z-19}{3z^3}\ ,\quad m_4^6=\frac1{z^2}\ ,\quad
m_4^4=\frac1{3z}\ .
\end{align}
In this limit the flow equations become
\begin{align}
\label{V15}
\partial_tu'=&\,-3u'+\tilde\chi
u''-\frac1{16\pi^2z}\bigg\{\frac{z'}{z}+\frac{(\ln z-1)u'''}{z}\bigg\}\ ,\nn\\
\partial_tz=&\,\tilde\chi z'-\frac1{32\pi^2z}\bigg\{2(\ln
z-1)\frac{z''}{z}-\bigg(5\ln z-\frac{43}{6}\bigg)\bigg(\frac{z'}{z}\bigg)^2\nn\\
&\qquad+\frac13(u''')^2-2u'''\frac{z'}{z}\bigg\}\ .
\end{align}

In order to include a possible singular behavior for $z\to\infty$ we rewrite
these equations in terms of the variable
\bel{V16}
s=\frac1z\ ,
\ee
resulting in
\begin{align}
\label{V17}
\partial_tu'=&\,-3u'+\tilde\chi u''+\frac1{16\pi^2}\big\{s'+(\ln
s+1)s^2u'''\big\}\ ,\nn\\
\partial_ts'=&\,\tilde\chi s'+\frac1{32\pi^2}\bigg\{2(\ln s+1)s^2s''+\bigg(\ln
s+\frac{19}{6}\bigg)s(s')^2\nn\\
&\qquad+\frac{s^3}{3}(u''')^2+2s^2s'u'''\bigg\}\ .
\end{align}
The region of interest concerns $s\ll1$.

\subsection*{Scaling solutions}

For scaling solutions the l.h.s. of the flow equation~\eqref{V17} vanishes,
$\partial_tu'=0$, $\partial_ts'=0$. There are two simple solutions. The first
corresponds to a free massless scalar field with constant $s$,
\bel{V18}
u'=0\ ,\quad s=s_0\ ,
\ee
corresponding to eq.~\eqref{V13A}. The second corresponds to
\bel{V19}
s=0\ ,\quad u=u_0+\frac{\lambda}{8}\tilde\chi^4 .
\ee
This corresponds to a massless scalar field with a quartic interaction. The
fluctuation effects are switched off due to perfect kinetic decoupling for
$s=0$. In view of the diverging kinetial the renormalized quartic coupling
$\lambda_R\sim\lambda/z^2=\lambda s^2$ vanishes.

We are interested in possible scaling solutions which interpolate between the
fixed points~\eqref{V18} and~\eqref{V19}. Let us first consider a region of
$\chi$ around some small $\bar\chi$ for which $s(\bar\chi)$ is also small. We
approximate $\ln s$ by a constant $\ln\bar s$, $\bar s=s(\bar\chi)$. In this
approximation we obtain the scaling solution
\bel{V20}
s=a\tilde\chi\ ,\quad
u=u_0+\frac{a\tilde\chi}{48\pi^2}+\frac{\lambda\tilde\chi^4}{8}\ ,
\ee
where
\bel{V21}
a^2=-\frac{32\pi^2}{\ln\bar s+19/6}\ .
\ee
This approximation requires small enough $\bar s$ such that $\ln\bar s+19/6<0$,
$a^2>0$. For positive $\tilde\chi$ we take $a>0$, for negative $\tilde\chi$ one
chooses $a<0$. Then $s$ is always positive, while the approximate solution is
not analytic at $\tilde\chi=0$. Besides the approximation for $\ln s$ we have
neglected terms which lead to relative corrections $\sim
a^2\lambda\tilde\chi^3$.

This approximate solution motivates for $\tilde\chi\to0$ the ansatz
\bel{V22}
s=4\pi\tilde\chi g\ ,
\ee
for which the scaling solution has to obey the differential equations
\bel{V23}
u'=\frac{g+\tilde\chi g'}{12\pi}+\frac{\tilde\chi}{3}\bigg[u''+\tilde\chi
g^2\gl\ln(4\pi\tilde\chi)+\ln g+1\gr u'''\bigg]\ ,
\ee
and
\begin{align}
\label{V24}
g+\tilde\chi g'=&-\frac12g(g+\tilde\chi g')^2\bigg(\ln(4\pi\tilde\chi)+\ln
g+\frac{19}{6}\bigg)\nn\\
&-\tilde\chi g^2(2g'+\tilde\chi g'')(\ln(4\pi\tilde\chi)+\ln g+1)\nn\\
&-\tilde\chi g^2(g+\tilde\chi g')u'''-\frac16\tilde\chi^2g^3(u''')^2\ .
\end{align}
It is not clear at the present stage if a satisfactory solution with smoothly
varying $g$ exists. We also do not know to what extent such a solution would
depend on the precise choice of the cutoff function.

The general issue for solutions with $s$ non analytic for $\tilde\chi\to0$
concerns the term $s'/(16\pi^2)$ in the flow equation~\eqref{V17} for $u'$. Even
a rather mild divergence of $s'$ (e.g. logarithmic) typically induces a similar
divergence in $u'$. Such a divergence results in a much stronger divergence of
$u'''$, which, in turn, can strongly affect the evolution of $s$. If this effect
is too strong the leading kinetial approximation will fail.

For a possible scaling $z\sim\tilde\chi^{-\alpha}$, $s\sim\tilde\chi^{\alpha}$
as discussed in appendix~\ref{app:FEK} the leading term for a possible scaling
solution obeys
\bel{B26}
u'-\frac13\tilde\chi u''=\frac{s'}{48\pi^2}\ .
\ee
For $s=48\pi^2c\tilde\chi^\alpha$ one finds
\bel{B27}
u'=\frac{3c\alpha}{4-\alpha}\tilde\chi^{\alpha-1}\ ,
\ee
and therefore
\bel{B28}
u'''=\frac{3c\alpha(\alpha-1)(\alpha-2)}{4-\alpha}\tilde\chi^{\alpha-3}\ .
\ee
This leads to contributions on the r.h.s. proportional to
$\tilde\chi^{5\alpha-6}$ or $\tilde\chi^{4(\alpha-1)}$, unless $\alpha=1$ or
$\alpha=2$. For $\alpha=2/3$ as discussed in appendix~\ref{app:FEK} for a
field-independent cutoff these terms diverge and overwhelm the terms retained in
the leading kinetial approximation. One concludes that the leading kinetial
approximation is not valid for a field-independent cutoff. This issue would
change for the particular case $z\sim\tilde\chi^{-2}$, $s\sim\tilde\chi^{2}$,
$\alpha=2$ for which one obtains $u'\sim\tilde\chi$, $u'''=0$.

\subsection*{Field dependent cutoff}

The situation is different for a suitable field-dependent cutoff. If one employs
a cutoff function proportional to the kinetial $K(\chi)$ the flow equation for
the effective potential becomes
\bel{B29}
\partial_tU=\frac12\int_	q\partial_t\gl
K(\chi)k^2r\gr\gl\Gamma^{(2)}+K(\chi)k^2r\gr^{-1}\ .
\ee
For a truncation in first order of a derivative expansion one has for constant
$\chi$
\bel{B30}
\Gamma^{(2)}=K(\chi)q^2+\partial_\chi^2U(\chi)\ .
\ee
With $\eta=-\partial_t\ln K(\chi)$ this yields
\bel{B31}
\partial_tU=\int_q\left(1-\frac\eta2+\frac12\partial_t\ln r\right)\gl
q^2+K^{-1}\partial_\chi^2U\gr^{-1}\ .
\ee
For $\partial_\chi^2U=0$ and $\partial_\chi\eta=0$ the flow of $U$ vanishes up
to a $\chi$-independent constant, $\partial_t\partial_\chi U=0$. A flat
potential remains flat for all $K$ which are proportional to a power of $k$,
\bel{B32}
K=f(\chi)k^{-\eta}\ .
\ee
In this case the leading kinetial approximation may be a reasonable
approximation.

The fixed point solution~\eqref{eq:AA} obeys both the flow equation for $K$ and
$U$. The flow equation~\eqref{FG17} remains valid, admitting also the scaling
solution~\eqref{FG26} with $\eta=-2$. For both scaling solutions the effective
action is invariant under the rescaling of $\chi$ by a constant factor $\beta$.
They realize a version of quantum scale symmetry. This version of quantum scale
symmetry, for which the metric (or the coordinates) are kept fixed, indeed
requires a flat potential. It is respected by the field-dependent cutoff, but
not by a field-independent cutoff. This could explain why no equivalent scaling
solution is found for the field-independent cutoff.

The possible scaling solutions,
\bel{B33}
\Gamma_k=\int_x\left\{\frac{\kappa^2}{2\chi^2}
\partial^\mu\chi\partial_\mu\chi+u_0k^4\right\}\ ,
\ee
with $\kappa=\kappa_0$ or $\kappa=ck^2$, describe a free massless composite
scalar field. It may not be equivalent to the trivial scaling solution where
$\kappa/\chi^2$ is replaced by one. It could define a new fixed point. For
example, there could be \qq{crossover scaling solutions} interpolating as a
function of $\tilde\chi$ from eq.~\eqref{B33} for $\tilde\chi\to0$ to the
trivial scaling solution for $\tilde\chi\to\infty$.

The simplified flow equation~\cite{CWSFE} employs a field-dependent cutoff. In
this approach one finds the scaling solution~\eqref{eq:AA}. The corrections due
to the field-dependence of the cutoff can be computed and are found to remain
consistent with the scaling solution~\eqref{eq:AA}.

\section{Free composite scalar fields}\label{app:FCSF}

In this appendix we employ microscopic field transformations with a linear
RG-kernel in order to discuss under which conditions a microscopic model with a
non-quadratic classical action can describe a free composite scalar field. The
issue concerns the question if one can find a microscopic realization of the
scaling solution~\eqref{eq:AA} or~\eqref{B33}.

\subsection*{Simple microscopic field transformations}

There are particular choices of microscopic variable transformations for which
the RG-kernel remains simple and the flow equation keeps its closed form.
Consider a transformation of the type
\bel{C9}
\vp'_k(x)=kf\left(\frac{\chi'(x)}{k}\right)=kf\gl\tilde\chi'(x)\gr\ ,
\ee
which is adapted to scale invariant scalar fields $\tilde\chi'=\chi'/k$ in four
dimensions. It implies
\bel{C10}
\partial_t\vp'_k(x)=k\partial_k\vp'_k(x)=\left(1-\frac{\partial\ln
f}{\partial\ln\tilde\chi}(x)\right)\vp'_k(x)\ .
\ee
For certain forms of $f$ one finds a simple behavior for $\partial_t\vp'_k(x)$,
as 
\bel{C11}
\vp'_k=ck\tilde\chi^{\prime\alpha}\ ,\quad
\partial_t\vp'_k(x)=(1-\alpha)\vp'_k(x)\ , \gamma_k=(1-\alpha)\vp\ .
\ee
This results in the flow equation
\begin{align}
\label{C12}
\partial_t\Gamma_{\vp,k}=&\,\frac12\tr\Big\{\gl\partial_t\cR_k+(2-2\alpha)\cR_k\gr\gl\Gamma_{\vp,k}^{(2)}+\cR_k\gr^{-1}\Big\}\nn\\
&-(1-\alpha)\int_x\frac{\partial\Gamma_{\vp,k}}{\partial\vp(x)}\vp(x)\ .
\end{align}

Another simple choice is
\bel{C13}
\vp'_k=ck\ln\tilde\chi'\ ,\quad \partial_t\vp'_k(x)=\vp'_k(x)-ck\ ,
\ee
with flow equation
\begin{align}
\label{C14}
\partial_t\Gamma_{\vp,k}=&\,\frac12\tr\Big\{\gl\partial_t\cR_k+2\cR_k\gr\gl\Gamma_{\vp,k}^{(2)}+\cR_k\gr^{-1}\nn\\
&-\int_x\frac{\partial\Gamma_{\vp,k}}{\partial\vp(x)}\gl\vp(x)-ck\gr\ .
\end{align}
For these choices the flow equation keeps a simple closed form. This can be
useful for approximations valid for certain ranges of the field.
The flow equations~\eqref{C12} and~\eqref{C14} are particular cases of the more
general setting of linear RG-kernels that lead to eq.~\eqref{C24}.
\subsection*{Free composite scalar fields}

A particular solution of the flow
equations~\eqref{C24},~\eqref{C12},~\eqref{C14} is the effective action for a
free massless scalar field,
\bel{C28}
\Gamma_{\vp,k}=\frac{Z(k)}{2}\int_x\partial^\mu\vp\partial_\mu\vp\ .
\ee
The first term in the flow equation~\eqref{C24} contributes only a
$\vp$-independent constant that may be neglected, and we conclude
\bel{C29}
\partial_t\Gamma_{\vp,k}=-\int_xZ(k)a(k)\partial^\mu\vp\partial_\mu\vp\ .
\ee
Comparison of coefficients yields an anomalous dimension
\bel{C30}
\partial_tZ=-2aZ\ ,\quad Z=Z_0\left(\frac{k}{k_0}\right)^{-2a}\ ,\quad
\eta=-\partial_t\ln Z=2a\ ,
\ee
with $a=1-\alpha$ and $\alpha=0$ for eq.~\eqref{C14}. (This can be extended to
free massive particles by adding to $\Gamma_{\vp,k}$ a $k$-dependent mass term,
and possibly also a term linear in $\vp$.)

If for large $k=\Lambda$ the microscopic effective average action
$\Gamma_{\vp,\Lambda}[\chi']$ can be brought to the form~\eqref{C28} (with a
possible additional mass term) by a suitable microscopic field
transformation~\eqref{C20}~\eqref{C22}, this form is preserved by the flow. This
defines a family of models that correspond to free composite scalar fields. For
a model formulated in terms of the canonical field $\chi'$, as for our
setting~\eqref{M1}, the issue concerns the initial value
$\Gamma_{\vp,\Lambda}[\chi]$. We found in sect.~\ref{sec:TFF} that for
$\Lambda\to\infty$ one has $\Gamma_{\chi,\Lambda}[\chi]=S[\chi]$, while
$\Gamma_{\vp,\Lambda}[\chi]$ will involve additional terms reflecting the
Jacobian of the microscopic field transformation. For a non-linear
kinetial~\eqref{M1} the classical action $S[\chi]$ needs additional terms,
typically a scalar potential, in order to achieve the initial value~\eqref{C9}
for $\Gamma_{\vp,\Lambda}$ despite the presence of a Jacobian. In contrast, a
measure $\sim\text{d}\chi'/\chi'$, which is linear in $\vp'$ for the
transformation~\eqref{FG5}, leads directly to an initial value of the
form~\eqref{C28}.

\subsection*{Conditions for linear RG-kernels}

We next discuss a few properties of the condition~\eqref{C22} for linear
RG-kernels. The \qq{linearity condition}~\eqref{C22} imposes a simple constraint
on
the scale dependence of $\tilde N$,
\begin{align}
\label{138A}
\partial_t\tilde
N=&\,\partial_t\frac{\partial\vp'}{\partial\chi'}=\left(1+\frac{\partial}{\partial\ln\chi'}\right)\partial_tN\nn\\
=&\,\left(1+\frac{\partial}{\partial\ln\chi'}\right)\left(a(k)N+\frac{b(k)}{\chi'}\right)=a(k)\tilde
N\ .
\end{align}
The general solution reads
\bel{138B}
\tilde N=\left(\frac{Z(k)}{Z_0}\right)^{-\frac12}\hat N(\chi')\ ,\quad
\eta(k)=-\partial_t\ln Z(k)=2a(k)\ ,
\ee
with $\hat N(\chi')$ independent of $k$. We conclude that the general form of
the field transformation which leads to a linear RG-kernel reads
\bel{140A}
\vp'=\left(\frac{Z(k)}{Z_0}\right)^{-1/2}\int_{\chi_0}^{\chi'}\text{d}\chi''\,\hat
N(\chi'')+f(k)\ .
\ee

Choosing $Z(k)=Z_0$, $f(k)=0$, $\hat N(\chi')=K^{1/2}(\chi')$ one can bring the
microscopic action~\eqref{M1} into the form for a free field $\vp'$, provided
$K(\chi')$ is positive and obeys some monotony properties. For a model with
classical action~\eqref{M1} the microscopic effective average action
$\Gamma_{\vp,\Lambda}[\vp]$ differs, however, from the microscopic or classical
action $S[\chi'(\vp)]$. Additional potential terms in $S[\chi']$ are necessary
for a microscopic effective action of the form~\eqref{C28}.

Eq.~\eqref{C22} admits logarithmic solutions,
\bel{C33}
\vp'=c(k)\ln\left(\frac{\chi'}{k}\right)\ ,\quad a=\partial_t\ln c\ ,\quad b=-c\
.
\ee
With 
\bel{C34}
\tilde N(\chi')=\frac{\partial\vp'}{\partial\chi'}=\frac{c(k)}{\chi'}\ ,\quad
K(\chi')=Z(k)\tilde N^2(\chi')\ ,
\ee
this maps
\bel{C35}
K(\chi')\partial^\mu\chi'\partial_\mu\chi'=Z(k)\frac{c^2(k)}{\chi^{\prime2}}
\partial^\mu\chi'\partial_\mu\chi'=Z(k)\partial^\mu\vp'\partial_\mu\vp'\ .
\ee
The flow of $Z$ according to eq.~\eqref{C30} becomes
\bel{36}
\partial_tZ=-2\partial_t\ln(c)Z\ .
\ee
In particular, for a powerlaw behavior of $c(k)$ the anomalous dimension is
constant
\bel{C37}
c(k)=c_0\left(\frac{k}{k_0}\right)^a\ ,\quad \partial_tZ=-2aZ\ ,\quad \eta=2a\ .
\ee

A particularly simple setting obtains by the limit $a\to0$ of eq.~\eqref{C33} by
choosing $c$ independent of $k$, such that $a=0$, $\eta=0$, and
\bel{C38}
\Gamma_{\vp,k}=\frac12\int_x\frac{Z_0c_0^2}{\chi^2}\partial^\mu\chi\partial_\mu\chi=\frac12\int_xZ_0\partial^\mu\vp\partial_\mu\vp\
.
\ee
For the particular field transformation,
\bel{C39}
\vp'=c_0\ln\left(\frac{\chi'}{k}\right)\ ,
\ee
the correction term in the flow equation vanishes (except for an irrelevant
constant) despite the $k$-dependence of the field transformation.

\subsection*{Scale symmetry}

The quantum effective action~\eqref{C38} realizes quantum scale
symmetry~\cite{CWQSS}. It is invariant under the scale- or dilatation
transformations
\bel{C40}
\chi(x)\to\beta\chi(x)\ ,\quad \vp(x)\to\vp(x)+\beta\ .
\ee
In this simple case quantum scale symmetry can be related to classical scale
symmetry. We can start with a scale invariant classical action
\bel{C41}
\Shat[\chi']=\frac12\int_x\frac{Z_0c_0^2}{\chi^{\prime2}}
\partial^\mu\chi'\partial_\mu\chi'\ ,
\ee
and a scale invariant measure
\bel{C42}
\int\cD\chi'=\prod_x\int_{-\infty}^{\infty}\frac{\text{d}\chi'(x)}{\chi'(x)}\ .
\ee
This is related to the formulation~\eqref{M2} with measure~\eqref{M4} by an
action
\bel{C43}
S[\chi']=\Shat[\chi']+\ln J[\chi']\ ,
\ee
with $J[\chi']$ the Jacobian~\eqref{M11}.

With
\bel{C44}
\frac{\text{d}\chi'}{\chi'}=\frac{\text{d}\vp'}{c_0}
\ee
the Jacobian of the field transformation to $\vp'$ does not involve the scale
$k$. It absorbs the term $\ln J[\chi']$ such that the microscopic or classical
action for $\vp'$ is the one for a free massless field. Up to a multiplicative
constant in the partition function or an additive constant in the effective
action, this setting describes a free massless scalar field theory for the field
$\vp'$, with measure $\prod_x\int\text{d}\vp'(x)$ and classical action
\bel{C45}
\Shat[\vp']=\frac{Z_0}{2}\int_x\partial^\mu\vp'\partial_\mu\vp'\ .
\ee
The difference to the action $\Sbar[\vp']$ in eq.~\eqref{M12} results from the
different measure~\eqref{C42}.

In this simple setting there are no field-dependent fluctuation effects. Up to
an additive constant this results for $k=0$ directly in the effective
action~\eqref{C38}. We observe that the source term is invariant under the
simultaneous shift $\vp'(x)\to\vp'(x)+\beta$, $\vp(x)\to\vp(x)+\beta$, which
translates classical scale symmetry to quantum scale symmetry. An infrared
cutoff term quadratic in $\vp'$ effectively maintains scale symmetry, if it can
be written in the form
\bel{C46}
\Delta_kS[\vp']=\frac12\int_x\partial^\mu\vp'\hat
r_k\left(-\frac{\partial^\mu\partial_\mu}{k^2}\right)\partial_\mu\vp'\ .
\ee
The issue is somewhat subtle if $\hat r_k(y)$ diverges for $y\to0$. For example,
$\hat r_k=y^{-1}$ results in a mass term $k^2$ in the classical action
supplemented by the cutoff, with a similar term in $\Gamma_{\vp,k}'[\vp]$. This
seems to indicate a violation of scale symmetry by the cutoff. The mass term is
subtracted, however, by the definition of $\Gamma_{\vp,k}[\vp]$, explaining the
$k$-independence of $\Gamma_{\vp,k}$ also for $k>0$. For the
measure~\eqref{C42},~\eqref{C44} we can write the functional integral defining
$\Gamma_{\vp,k}$ in terms of $\hat\vp(x)=\vp'(x)-\vp(x)$ which is invariant
under the simultaneous shift in $\vp'$ and $\vp$.

For a field-independent cutoff the definition of $\Gamma_{\chi,k}$ violates
scale symmetry by the cutoff quadratic in $\chi'$. In contrast, the formal
definition of the quantum effective action $\Gamma_\chi[\chi]$ $(k=0)$ preserves
scale symmetry if $S[\chi']$ is scale invariant. The source term is invariant
under a simultaneous scaling of $\chi'$ and $\chi$. The Jacobian for global
rescalings of the linear measure~\eqref{M4} is a field independent factor which
only adds an irrelevant additive constant to $\Gamma_\chi$. One may therefore
ask if there exists an appropriate setting for the flow equation that maintains
scale symmetry in case of the measure~\eqref{M4} for scale invariant $S$. For
the measure~\eqref{M4} the transition to $\Gamma_\vp$ violates scale symmetry by
the Jacobian of the field transformation. Then $\Gamma_\vp$ is not expected to
be scale symmetric. For $\Gamma_\chi$ one may imagine a cutoff term involving
the macroscopic field $\chi$ in the form $\cR_k\sim\chi^{-2}$. In this setting
the cutoff is invariant under simultaneous scale transformations of $\chi'$ and
$\chi$. We discuss this case of a field-dependent cutoff in
appendices~\ref{app:FEK},~\ref{app:FKP}, where one finds indeed a scaling
solution with scale symmetry. 

\bigskip
\noindent

\nocite{*} 
\bibliography{refs}
\end{document}